\documentclass[prd,twocolumn,showpacs,preprintnumbers,floats,floatfix,nofootinbib]{revtex4}

\usepackage{graphicx}
\usepackage{dcolumn}

\begin{document}
\title{Bouncing Braneworld Cosmologies and Initial Conditions to Inflation}
\author{R. Maier and I. Dami\~ao Soares}
\hspace{0.5cm}

\affiliation{Centro Brasileiro de Pesquisas F\'\i sicas, Rua
Dr. Xavier Sigaud 150, Urca
\\Rio de Janeiro. CEP 22290-180-RJ, Brazil}
\author{E. V. Tonini}

\affiliation{Centro Federal de Educa\c c\~ao Tecnol\'ogica do
Esp\'\i rito Santo\\ Avenida Vit\'oria, 1729, Jucutuquara, Vit\'oria
CEP 29040-780-ES, Brazil}

\date{\today}

\begin{abstract}

We examine the full nonlinear dynamics of closed FRW universes in the framework of
D-branes formalism. Friedmann equations contain additional terms arising from the
bulk-brane interaction that provide a concrete model for nonsingular bounces
in the early phase of the universe. We construct nonsingular cosmological scenarios
sourced with perfect fluids and a massive inflaton field which are past eternal,
oscillory and may emerge into an inflationary
phase due to nonlinear resonance mechanisms. Oscillatory behaviour becomes
metastable when the system is driven into a resonance window
of the parameter space of the models, with consequent break-up of KAM tori that trap the
inflaton, leading the universe to the inflationary regime.
A construction of the resonance chart of the models is made.
Resonance windows are labeled by an integer $n \geq 2$, where $n$ is related to the ratio
of the frequencies in the scale factor/scalar field degrees of freedom.
They are typically small compared to volume of the whole parameter
space, and we examine the constraints imposed by nonlinear resonance
in the physical domain of initial configurations so that
inflation may be realized. We discuss the complex dynamics arising in this
pre-inflationary stage, the structural stability of the resonance pattern and some
of its possible imprints in the physics of inflation. We also approach the
issue of initial configurations that are connected to a chaotic exit to inflation.
Pure scalar field bouncing cosmologies are constructed. Contrary to models with
perfect fluid components, the structure of the bouncing dynamics is highly
sensitive to the initial amplitude and to the mass of the inflaton;
dynamical potential barriers allowing for bounces
appear as a new feature of the dynamics.
We argue that if our actual Universe is a brane inflated by a parametric resonance
mechanism triggered by the inflaton, some observable cosmological parameters
should then have a signature of the particular resonance from which the brane inflated.
\end{abstract}

\label{PACS number:}
\maketitle

\section{Introduction}

The issue of the initial conditions of our present Universe
is connected to the problem of the initial singularity
and to the possible solutions adopted to circumvent this problem,
which lie in the realm of a quantum theory of gravitation.
In fact we may consider that the initial conditions of our present
expanding Universe were fixed when the early Universe emerged from
a Planckian regime and started its classical evolution. However by
evolving back the initial conditions using Einstein classical equations
the Universe is driven towards a singular point where the classical regime is
no longer valid. This is an indication that classical General Relativity
is not a complete theory and in this domain quantum processes must be taken into
account. Therefore initial conditions from which our classical universe evolved
should crucially depend on the version of quantum gravity theory adopted to describe
the dynamics in the neighborhood of the classical singularity. This implies that
recent observational results in Cosmology could in principle guide us
in narrowing the possibilities of choices. Inflationary
cosmology, for instance, although a highly appealing theoretical paradigm,
relies on assumptions about how the universe emerged from the cosmic singularity. In this vein
models for a preinflationary phase, including quantum corrections and consistent
with the inflationary paradigm, are important to be examined.
\par Among several propositions to describe the dynamics in this preinflationary
semiclassical domain are, for instance, quantum loop cosmology\cite{reza}
and the string based formalism of $D$-branes\cite{string}, both of them leading to
corrections in Einstein's equations and encompassing General
Relativity as a classical (low energy) limit. In the case of spatially
homogeneous and isotropic cosmologies, the basic resulting distinction
between the two approaches lies in the corrections introduced in
Friedmann's Hamiltonian constraint: quantum loop cosmology leads
to corrections in the kinetic energy term of matter fields while bulk-brane corrections
lead to extra potential energy terms. In both cases we may have bounces in the
scale factor corresponding to the avoidance of a singularity in the models.
\par In the present paper we adhere to the string based formalism of $D$-branes.
In this scenario extra dimensions are introduced, the bulk space,
and all the matter in the universe would be trapped on a brane with
three spatial dimensions; only gravitons would be allowed to leave
the surface and move in the full bulk\cite{RS}.
At low energies General Relativity is recovered but at high energies significant
changes are introduced in the gravitational dynamics. Our main interest here
is connected to the high energy/quantum corrections that are dominant in
the neighborhood of the singularity, resulting in a repulsive force that
avoids the singularity and leads the universe to undergo nonsingular bounces.
An elegant geometrical derivation of braneworld dynamics embedded in 5-dim spacetimes
may be found in Refs. \cite{maeda}-\cite{maartens} where both
high-energy local corrections as well as
nonlocal bulk corrections on a Friedmann-Robertson-Walker (FRW) brane are analyzed.
Bouncing braneworld models were constructed by Shtanov and Sahni\cite{sahni}
based upon a Randall-Sundrum type action with one extra timelike dimension.
\par We here examine the full nonlinear dynamics of spatially closed FRW preinflationary
braneworld models with a massive scalar field (the inflaton) and
several noninteracting perfect fluids. The matter fields evolve on the brane,
where high-energy/quantum gravity corrections due to the bulk are included and
implement nonsingular bounces. We have previously approached
analogous models with a radiation fluid plus a scalar field
in the form of small perturbations, where the brane corrections
were due to the radiation fluid only or to a phantom-type fluid\cite{aranha}.
\par With the full nonlinear dynamics new possibilities for cosmological scenarios
arise, as nonsingular oscillatory bouncing cosmologies sourced with a pure scalar field,
or with a scalar field plus several perfect fluid components that allow to
model the effect of dark matter together with barionic matter in the
gravitational dynamics. Such nonsingular oscillatory solutions have the
theoretical advantage of avoiding the problem of initial conditions at
past infinity occurring with one-single bounce solutions and, further,
are favored by entropy considerations.
We make a detailed examination of nonlinear parametric resonance mechanisms that
are present in the full dynamics and turn these bounded oscillatory solutions metastable
allowing the model to emerge naturally into an inflationary phase.
We consider the restriction such mechanisms impose on the initial configurations
so that the models may realize inflation.
\par We organize the paper as follows. In Section II we
give a brief introduction to the framework of the brane-world formulation,
making explicit the assumptions used in obtaining the dynamics,
and derive the full dynamical equations of the models. In Section III
we describe some basic structures (as critical points,
invariant planes and attractors at infinity) that constitute the skeleton
of phase space and allow to organize the dynamics in phase space.
In section IV we restrict the matter content of the models to
a massive scalar field plus dust and radiation, that constitute a minimal set
of ingredients appropriate for a preinflationary model, and analyze the
constraints on the parameters of the model so that the dynamics may allow for
bounded oscillatory bouncing solutions. In Sections V and VI we make a semi-analytical
approach to nonlinear resonance phenomena in the models, that may turn the oscillatory
bounded solutions into metastable ones, with an inflationary behavior,
and we construct the resonance charts of the dynamics.
In Section VII we treat the case of bouncing cosmologies sourced by a pure
scalar field. Section VIII is devoted to the exam of the dynamics of
initial condition sets connected to a chaotic exit to inflation.
Section IX contains the conclusions and final discussions.
Throughout the paper we use units such that $\hbar=c=1$.
\section{The Model and Its Dynamics}
Our task here is to derive the full nonlinear dynamical equations of the models,
(actually a 4-dim autonomous dynamical system with one
first integral) and to analyze structure of the associated phase space.
In the framework of D-brane formalism, we consider a closed
Friedmann-Robertson-Walker (FRW) metric on the 4-dim braneworld
embedded in a 5-dim conformally flat bulk. The matter content of the models
is constituted of a scalar field $\phi$ plus several noninteracting
perfect fluids, each with equation of state $p_i=\alpha_{i}\rho_i$.
These matter fields are constrained to propagate on the brane only.
\par We start by giving a brief introduction to brane-world theory,
making explicit the specific assumptions used in
obtaining the dynamics of the model. We rely on references
\cite{sahni}-\cite{maartens}, and our notation
basically follows \cite{wald}. Let us start with a 4-dim Lorentzian
brane $\Sigma$ with metric $g_{ab}$, embedded in a 5-dim conformally flat bulk $\cal{M}$ with
metric $g_{AB}$. Capital Latin indices range from 0 to 4, small Latin
indices range from 0 to 3. We regard $\Sigma$ as a common boundary of two pieces
${\cal{M}}_{1}$ and ${\cal{M}}_{2}$ of $\cal{M}$ and the metric $g_{ab}$ induced on the brane by the metric of the two pieces should coincide although the extrinsic
curvatures of $\sigma$ in ${\cal{M}}_{1}$ and ${\cal{M}}_{2}$ are allowed to be different.
The action for the theory has the general form
\begin{eqnarray}
\label{eq4r}
\nonumber
S&=&\frac{1}{2 \kappa_5} \Big[ \int_{{\cal{M}}_1}\Big( ~ ^{(5)}R - 2 \Lambda_5 \Big) + 2 \epsilon \int_{\Sigma} K_1 \\
\nonumber
&+& \int_{{\cal{M}}_2}\Big( ~ ^{(5)}R - 2 \Lambda_5 \Big)
- 2 \epsilon \int_{\Sigma} K_2 \Big] \\
&+& \frac{1}{2}\int_{\Sigma} \Big( \frac{1}{\kappa_4}~ ^{(4)}R - 2 \sigma \Big)+
\int_{\Sigma}L_4 (g_{ab}, \rho, \phi).
\end{eqnarray}
In the above $^{(5)}R$ is the Ricci scalar of the Lorentzian 5-dim metric $g_{AB}$
on $\cal{M}$, $\Lambda_5$ is the 5-dim cosmological constant
and $^{(4)}R$ is the scalar curvature of the induced metric $g_{ab}$
on $\Sigma$. The parameter $\sigma$ is denoted the brane tension,
and $\kappa_{5}$ and $\kappa_{4}$ are Einstein constants in five- and four-dimensions,
respectively. The unit vector $n^{A}$ normal to the boundary $\Sigma$ has
norm $\epsilon$. If $\epsilon=-1$ the signature of the bulk space is
$(-,-,+,+,+)$, so that the extra dimension is timelike.
The quantity $K=K_{ab}~ g^{ab}$ is the trace of the
symmetric tensor of extrinsic curvature
$K_{ab}= Y_{,a}~^{C}~Y_{,b}~^{D}~{\nabla_{C}}{n_{D}}$, where $Y^{A}(x^a)$ are the
embedding functions of $\Sigma$ in $\cal{M}$\cite{sahni}. Also

\begin{eqnarray}
L_{4}= \sqrt{- g} \Big[\sum_{i}\rho_{i}- \frac{1}{2}\Big(g^{ab}\phi_{,a} \phi_{,b}+ m^2
\phi^{2} \Big)- \frac{\xi}{2} ~ {^{(4)}R}\phi^{2}\Big]
\label{eq5r}
\end{eqnarray}
is the Lagrangean density of the four dimensional massive inflaton field $\phi$
plus the perfect fluids (with equation of state $p_i= \alpha_{i}\rho_{i}$),
whose dynamics is restricted to the brane $\Sigma$. They interact only with the
induced metric $g_{ab}$. We further assume that the inflaton field is nonminimally coupled
with $g_{ab}$, with coupling parameter $\xi$. All integrations over the bulk and the
brane are taken with the natural volume elements $\sqrt{-\epsilon {^{(5)}}g}~ d^{5}x$ and $\sqrt{- {^{(4)}}g}~ d^{4}x$ respectively.
\par Variations that leave the induced metric on $\Sigma$
intact result in the equations
\begin{eqnarray}
\label{eq6r}
^{(5)}G_{AB}+ \Lambda_5~g_{AB}=0,
\end{eqnarray}
while considering arbitrary variations of $g_{AB}$ and taking into account (\ref{eq6r})
we obtain
\begin{eqnarray}
\label{eq7r}
^{(4)}G_{ab}+\epsilon ~\frac{\kappa_4}{\kappa_5}\Big( S^{(1)}_{ab}-S^{(2)}_{ab}\Big )
=\kappa_4\Big(\tau_{ab}-\sigma g_{ab} \Big),
\end{eqnarray}
where $S_{ab} \equiv K_{ab}-K g_{ab}$. $\tau_{ab}$ is the energy-momentum tensor of the
matter fields on the brane, resulting from (\ref{eq5r}). In the limit
$\kappa_4 \rightarrow \infty$ equation (\ref{eq7r}) reduces to
the Israel-Darmois
junction condition\cite{israel}
\begin{eqnarray}
\label{eq9r}
\Big( S^{(1)}_{ab}-S^{(2)}_{ab}\Big )
=\epsilon~ \kappa_5\Big(\tau_{ab}-\sigma g_{ab} \Big).
\end{eqnarray}
We impose the $Z_2$-symmetry\cite{maartens} and use the junction conditions (\ref{eq9r})
to determine the extrinsic curvature on the brane,
\begin{eqnarray}
\label{eq10r}
K_{ab}=-\frac{\epsilon}{2} \kappa_5 \Big[(\tau_{ab}-\frac{1}{3}\tau g_{ab})+\frac{\sigma}{3} g_{ab} \Big].
\end{eqnarray}
Now using Gauss equation\\ $^{(4)}R_{abcd}=~^{(5)}R_{MNRS} Y^{M}_{,a} Y^{N}_{,b} Y^{R}_{,c} Y^{S}_{,d} +\epsilon \Big(K_{ac}K_{bd}-K_{ad}K_{bc} \Big)$ together with equations (\ref{eq6r})
and (\ref{eq10r}) we arrive at the induced field equations on the brane
\begin{eqnarray}
\label{eq11r}
\nonumber
^{(4)}G_{ab}=- \Big(\frac{\Lambda_{5}}{2}+ \frac{1}{12}~ \kappa_{5}^{2} ~\epsilon ~\sigma^2 \Big)~ g_{ab}\\
+\frac{\kappa_{5}^{2}}{6}~ \epsilon~\sigma ~ \tau_{ab}
+\epsilon~\kappa_{5}^2~\pi_{ab}~.
\end{eqnarray}
In the above
\begin{eqnarray}
\label{eq12r}
\pi_{ab}=-\frac{1}{4}\tau_{ac}\tau_{b}^{c}+\frac{1}{12}\tau\tau_{ab}+\frac{1}{8}~g_{ab}\tau_{cd} \tau^{cd}
-\frac{1}{24}~g_{ab}~\tau^{2}.~
\end{eqnarray}
We remark the absence of the conformal tensor projection in Eq. (\ref{eq11r})
since the 4-dim brane is embedded in a conformally flat bulk, which is the case of the FRW
brane to be considered. Accordingly Codazzi's equations imply that
\begin{eqnarray}
\label{eq11rr}
D_{a}\tau^{a}_{b}=0,
\end{eqnarray}
where $D_a$ is the covariant derivative with respect to the induced metric $g_{ab}$.
Together with the contracted Bianchi's identities in (\ref{eq11r}) it results $D_{a}\pi^{a}_{b}=0$,
which corresponds to a sufficient condition for $g_{ab}$ to be embedded in a conformally
flat bulk.
\par Equations (\ref{eq11r}) and (\ref{eq11rr}) are the dynamical equations of the
gravitational field on the brane. In Eq. (\ref{eq11r}) we identify
\begin{eqnarray}
\label{eq11rri}
\Lambda_{4}=\Big(\frac{\Lambda_{5}}{2}+\frac{1}{12}\kappa_{5}^{2} ~\epsilon ~\sigma^2 \Big),~~~~G_N=\frac{\kappa_5^{2} \sigma \epsilon}{48\pi},
\end{eqnarray}
respectively as the effective cosmological constant, and the effective
Newton's gravitational constant in the brane. Both depend basically on the
brane tension $\sigma$. Eq. (\ref{eq11r}) is similar to Einstein equations
in 4 dimensions, except by the second term in the RHS that is a correction resulting
from the brane-bulk interaction quadratic in the extrinsic curvature, while Eq. (\ref{eq11rr})
is the conservation law for the matter on the brane.
We recall that for the evaluation of the extrinsic curvature
(\ref{eq10r}) we use the energy-momentum of the matter fields
on the brane. In our model they are described by the
Lagrangean density (\ref{eq5r}).
\par Let us consider the FRW metric on the brane given by the line element
\begin{eqnarray}
\label{eq14r}
ds^2=-dt^2+a(t)^2\Big[\frac{dr^2}{1-kr^2}+r^2 (d\theta^2+ \sin^2\theta d\phi^2) \Big],~~
\end{eqnarray}
where $a(t)$ is the scale factor. For the noninteracting perfect fluids considered
in (\ref{eq5r}), each with equation of state $p_i=\alpha_{i}\rho_i$, we have then
\begin{eqnarray}
\label{eq15r}
\rho_i=\frac{E_{i}}{a^{3(1+\alpha_i)}},
\end{eqnarray}
where $E_i$ is a constant of motion. The components of the tensor $\pi_{ab}$ (cf. (\ref{eq12r})) are now expressed as
\begin{eqnarray}
\label{eq15r1}
\pi_{00}=\frac{1}{12} \Big \{\sum_{i} \rho_i+ \rho_{\phi}+ \Delta  \Big\}^2,
\end{eqnarray}
\begin{eqnarray}
\label{eq15r2}
\nonumber
\pi_{ij}=\frac{1}{12}\Big \{\sum_{i} \rho_i&+& \rho_{\phi}+ \Delta \Big\}^2 g_{ij} +
\frac{1}{6}\Big \{\sum_{i} \rho_i+ \rho_{\phi}+ \Delta \Big\} \times\\
\nonumber
\Big \{ \sum_{i} p_i&+& p_{\phi} +2 \xi \Big[ (m^2 +\xi R)\phi^2-{\dot{\phi}}^2 +
{\frac{\dot{a}}{a}}~\phi {\dot{\phi}} \Big]\\
&-&\xi \phi^2(2\frac{\ddot{a}}{a}+{\frac{{\dot{a}}^2}{a^2}}+\frac{k}{a^2})\Big\} g_{ij},
\end{eqnarray}
where
\begin{eqnarray}
\label{eq15r3}
\rho_{\phi}=\frac{1}{2} \Big ( {\dot{\phi}}^2+ m^2 \phi^2 \Big),~~~~p_{\phi}=\frac{1}{2} \Big ( {\dot{\phi}}^2- m^2 \phi^2  \Big),
\end{eqnarray}
\begin{eqnarray}
\label{eq15r4}
\Delta=3 \xi\Big[\Big({\frac{{\dot{a}}^2}{a^2}}+\frac{k}{a^2}\Big)\phi^2+2 {\frac{\dot{a}}{a}}\phi \dot{\phi} \Big],
\end{eqnarray}
and the curvature scalar $R=6\Big[\ddot{a}/a+(\dot{a}/a)^2+k/a^2 \Big]$.
The equations of motion (\ref{eq11r}) and (\ref{eq11rr})
reduce then to
\begin{eqnarray}
\label{eq16r}
\nonumber
\Big(\frac{\dot{a}}{a}\Big)^2+\frac{k}{a^2}-\frac{\Lambda_4}{3}=\frac{8 \pi G_{N}}{3}\Big[ \sum_{i}\rho_{i} +\frac{1}{2} {\dot{\phi}}^2+\frac{1}{2}m^2 {\phi}^2\\
\nonumber
+3 \xi \Big( (\frac{\dot{a}}{a})^2 +\frac{k}{a^2}\Big)\phi^2+6\xi \frac{\dot{a}}{a}\phi \dot{\phi}\Big]+
\frac{\epsilon {\kappa}_{5}^{2}}{36}\Big[\sum_{i}\rho_{i} +\frac{1}{2} {\dot{\phi}}^2\\+\frac{1}{2}m^2 {\phi}^2
+3 \xi \Big( (\frac{\dot{a}}{a})^2 +\frac{k}{a^2}\Big)\phi^2+6\xi \frac{\dot{a}}{a}\phi \dot{\phi} \Big]^2,~~
\end{eqnarray}
\begin{eqnarray}
\label{eq17r}
\ddot{\phi}+3\frac{\dot{a}}{a}\dot{\phi}+m^2 \phi+6 \xi \Big[ \frac{\ddot{a}}{a}+(\frac{\dot{a}}{a})^2+\frac{k}{a^2}\Big]=0,
\end{eqnarray}
respectively. Eq. (\ref{eq17r}) is the Klein-Gordon (KG) equation for the
inflaton field $\phi$ while Eq. (\ref{eq16r}) is the $(0,0)$ equation in
(\ref{eq11r}). Eq. (\ref{eq16r}) is in fact a first
integral of the $(i,j)$ equations provided the KG equation holds.
\par The nonminimal coupling of the inflaton with gravitation
considered here is partly motivated by quantum calculations in curved
spacetimes (taking into account quantum backreaction, renormalization, etc.)
and partly by enlarging the possibilities of constructing successful inflationary and
preinflationary scenarios(cf. for instance \cite{nonminimal}). The case $\xi=0$ is the usual minimal
coupling of the scalar field with gravitation, and $\xi=1/6$ is the
so-called conformal coupling\cite{birrel}.
\par It is not difficult to see from Eq. (\ref{eq16r}) that the choice $\epsilon=-1$
(corresponding to the extra dimension with timelike character) will
implement nonsingular bounces in the dynamics of the scale factor $a$,
implying that the models are nonsingular. In this case the brane tension
$\sigma$ is required to be negative in order that the subsequent evolution of the
universe be compatible with observations (cf. Eq. (\ref{eq11rri})).
\par In the remaining of the paper we will restrict ourselves to the
case $\epsilon=-1$ and positively curved FRW universes ($k>0$), as we are interested
in preinflationary nonsigular dynamics with metastable oscillatory behaviour.
Partly for analytical and numerical simplicity we will also restrict our
analysis to the case of conformal coupling $\xi=1/6$. The basic features of this case
encompass that of the minimal coupling case $\xi=0$. Arbitrary $\xi$'s correspond to
a higher dimensional parameter space, the scope of which is beyond the purpose
of the present paper. The case of pure scalar field bouncing cosmologies will demand a separate
treatment, and will be dealt with in Section VII. Finally, for numerical computation
purposes we fix $\kappa_{5}^{2}=6$ and $k=1$.
\par Within the above restrictions, using the conformal time variable $\tau$
(defined by $d \tau=dt/a$) and the rescaled scalar field $\varphi=a \phi$,
Eqs. (\ref{eq16r}) and (\ref{eq17r}) simplify to
\begin{eqnarray}
\label{eq18}
{{a}^{\prime}}^2+{a^2}-\frac{\Lambda_4}{3}~a^4=\frac{|\sigma|}{3}\Gamma(\rho_i,\varphi)
-\frac{1}{6a^4}\Big[\Gamma(\rho_i,\varphi) \Big]^2,~~
\end{eqnarray}
where
\begin{eqnarray}
\Gamma(\rho_i,\varphi) \equiv \Big[{a}^4 \sum_{i}\rho_{i} +\frac{1}{2}\Big( {{\varphi}^{\prime}}^2+(1+m^2 {a}^2) {\varphi}^2\Big) \Big],~~
\end{eqnarray}
and
\begin{eqnarray}
\label{eq19}
{\varphi}^{\prime\prime}+(1+m^2{a}^2)~ \varphi=0,
\end{eqnarray}
where a prime denotes derivative with respect to the conformal time $\tau$.
Eqs. (\ref{eq18}) and (\ref{eq19}) constitute the basic dynamical equations to be
dealt with in this paper, defined in the phase space
($a,\varphi,{a}^{\prime},{\varphi}^{\prime}$).
\par We note that for $m=0$ the dynamics is separable and, as a consequence, integrable.
The Klein-Gordon equation (\ref{eq19}) has the first integral ${\mathcal{E}}_{\varphi}^{0}=\Big({{\varphi}^{\prime}}^2+{\varphi}^2 \Big)/2$ inthis case.
\section{The skeleton of phase space}

From the above equations (\ref{eq18}) and (\ref{eq19}) we derive the dynamical system
\begin{eqnarray}
\nonumber
{\varphi}^{\prime}&=&p_{\varphi},\\
\nonumber
{p_{\varphi}}^{\prime}&=&-(1+m^2{a}^2)~ \varphi,\\
\nonumber
{a}^{\prime}&=&p_a/6,
\end{eqnarray}
\begin{widetext}
\begin{eqnarray}
\label{eq21}
\nonumber
{p_a}^{\prime}&=&-6 a+4 \Lambda_4 a^3- |\sigma|
\Big[\sum_{i}(3 \alpha_i-1)\frac{E_i}{a^{3\alpha_i}}-m^2 a {\varphi}^{2} \Big]
+\Big[ \sum_{i}\frac{E_i}{a^{3\alpha_i+1}} +\frac{1}{2 a^2}\Big( {{\varphi}^{\prime}}^2+(1+m^2 a^2)\varphi^2 \Big)\Big]\\
&\times& \Big[ \sum_{j}(3 \alpha_j+1)\frac{E_j}{a^{3\alpha_j+2}}+\frac{1}{a^3}\Big( {{\varphi}^{\prime}}^2+\varphi^2 \Big)\Big].
\end{eqnarray}
\end{widetext}
The first and third equations (\ref{eq21}) are mere redefinitions. ($p_a,a$) can be shown
to be canonically conjugated (cf. description of the invariant plane dynamics, for
instance). This is not the case of ($p_{\varphi},\varphi$), since the first
integral (\ref{eq18}) is not a Hamiltonian constraint in these variables due the
presence of fourth-order powers in ${{\varphi}^{\prime}}$.
\par Three basic structures organize the dynamics in the phase space of
the above dynamical system, namely, (i) an invariant plane, (ii) critical points and
(iii) separatrices, allowing us to give a global description of the motion of the models.
\par The invariant plane is defined by
\begin{eqnarray}
\label{eq22}
{\varphi}=0,~~p_{\varphi} \equiv {\varphi}^{\prime}=0,
\end{eqnarray}
where the dynamics is integrable; orbits with initial conditions in this plane
are totally contained in it, actually being similar to the dynamics in the sector
($a,p_a$) of the separable case. A first integral of motion is given by the Hamiltonian constraint
\begin{eqnarray}
\label{eq23}
\frac{{p_{a}}^2}{12}+V(a)-|\sigma|E_{\rm rad}=0
\end{eqnarray}
where the potential $V(a)$ is defined
\begin{eqnarray}
\label{eq24}
V(a)=3{a^2}-\Lambda_4 a^4 - |\sigma| \sum_{i \neq {\rm rad}}
\frac{E_i}{a^{3{\alpha}_i-1}}
+\frac{1}{2}\Big(\sum_{i}\frac{E_i}{a^{3{\alpha}_i+1}}\Big)^2.~~
\end{eqnarray}
The equations of motion are equivalent to Hamilton's equations generated from (\ref{eq23}),
corresponding to the third and fourth equations (\ref{eq21}) restricted to the
invariant plane (\ref{eq22}).
\par A careful examination shows that for perfect fluids with $ -1/3 < \alpha_i \leq 1$
the last term in the potential (\ref{eq24}) acts as an infinite potential barrier and
is responsible for the avoidance of the singularity $a=0$. These potential
corrections are equivalent to fluids with negative energy densities. This is
in accordance with the fact that indeed quantum effects can violate the classical energy
conditions, and may avoid curvature singularities where classical general relativity breaks down\cite{ford}. Such quantum violations tend to occur on short scales and/or at high curvatures, which is the case in our present models.
\par Critical points in the finite region of phase space are stationary solutions of
Eqs. (\ref{eq21}), namely, the points ($p_{\varphi}=0,\varphi=0,p_a=0,a=a_{\rm crit}$),
at which the RHS of Eqs. (\ref{eq21}) vanish. Obviously the critical points
are contained in the invariant plane, with $a=a_{\rm crit}$ being the
real positive roots of $V^{\prime}(a)=0$, where a prime here denotes derivative
with respect to $a$. Depending on the values of the parameters
($\Lambda_4, \sigma, E_i,\alpha_i$) we may have from one to several
real positive $a_{\rm crit}$. However it is not difficult to verify
that not all of them satisfy the constraint equation (\ref{eq18}) and
that at most two critical points are present, associated with one minimum
and one maximum of $V(a)$. As a matter of fact, this is
the case for a fixed $\Lambda_4$ and sufficiently bounded values of $E_i$.
These limiting conditions on the $E_i$ have not in general a closed analytical form
except for the cases of dust or radiation, and will be dealt with in next Section.
For $\Lambda_4=0$ the dynamical system has one critical point
only, corresponding to a minimum of the potential $V(a)$; such configurations
are not of interest for us since they corresponding to perpetually
nonsingular oscillating universes, a scenario where inflation cannot be realized.
\par To examine the nature of the critical points, and consequently the
nature of the motion in their neighborhood, we linearize the dynamical
system (\ref{eq21}) about each $a_{\rm crit}$. The resulting $4 \times 4$
constant matrix of the linearization has the four eigenvalues
\begin{eqnarray}
\label{eq25}
\lambda_{1,2}= \pm i \sqrt{1+m^2 a_{\rm crit}^{2}}~,~~~~\lambda_{3,4}=\pm \sqrt{-V^{\prime \prime}(a_{\rm crit})/6}~.
\end{eqnarray}
related to the linearized motion about $a_{\rm crit}$ in the
sector ($\varphi,p_{\varphi}$) and in the sector ($a,p_a$), respectively.
This characterizes the minimum of $V(a)$ as a center $P_0$, with two pairs of
complex conjugated imaginary eigenvalues ($V^{\prime \prime}(a_{\rm crit})>0$),
and the maximum $P_1$ as a saddle-center\cite{saddle} with one pair of real
eigenvalues with opposite signs ($V^{\prime \prime}(a_{\rm crit})<0$) and one
pair of imaginary eigenvalues. These results can be easily interpreted if we expand the integral
of motion (\ref{eq18}) about the critical points as
\begin{eqnarray}
\label{eq26s}
\nonumber
{\cal{H}}&\equiv& \frac{1}{12}p_a^{2}+\frac{1}{2} V^{\prime \prime}(a_{\rm crit})~ (a-a_{\rm crit})^2\\
\nonumber
&-&\frac{|\sigma|}{2}\Big(p_{\varphi}^{2}+(1+m^2 a_{\rm crit}^2)\varphi^2 \Big)\\
&+&E_{\rm crit}-|\sigma|E_{rad}+
{\cal{O}}(3)=0,
\end{eqnarray}
where ${\cal{O}}(3)$ denotes higher order terms in the expansion and $E_{{\rm crit}}\equiv V(a_{\rm crit})$ is the energy of the respective critical point. In a small neighborhood of
the critical point these higher-order terms can be neglected and the motion is separable into the two sectors with approximate constant of motions
\begin{eqnarray}
\label{eq26}
\nonumber
E_{(a)}&=& \frac{1}{12}p_a^{2}+\frac{1}{2} V^{\prime \prime}(a_{\rm crit})~ (a-a_{\rm crit})^2,\\
E_{(\varphi)}&=& \frac{1}{2}\Big(p_{\varphi}^{2}+(1+m^2 a_{\rm crit}^2)\varphi^2 \Big),
\end{eqnarray}
with $E_{(a)}-|\sigma|E_{(\varphi)}+ E_{\rm crit}-|\sigma|E_{rad} \sim 0$ and $|E_{\rm crit}-|\sigma|E_{rad}|$ small. It is immediate to see that the sector associated with
$E_{(\varphi)}$ always correspond to rotational motion in the variables ($\varphi,p_{\varphi}$) about the critical points, while the sector associated with $E_{(a)}$ corresponds
to either (i) rotational motion or (ii) hyperbolic motion in the variables ($a,p_a$)
about the  respective critical point; namely, the minimum of the potential $P_0$ corresponds
to a center and the maximum $P_1$ corresponds to a saddle-center, as mentioned before.
$P_0$ and $P_1$ define, respectively, a stable and an unstable
static Einstein universe. The stable Einstein universe configuration has no classical
analogue, arising from the dynamical balance between the perfect fluid energy
content and the negative energy density connected to the high energy/quantum
corrections in the potential. We are now ready to describe the topology of
the motion about the critical points and its extension to a nonlinear neighborhood
of these points. This will be done in next Section for the case of a preinflationary model containing cold dark matter (dust), dark energy (a positive cosmological constant)
and radiation.
\par Finally from the saddle-center critical point (when present) there emerges
a structure of separatrices $\mathcal{S}$ contained in the invariant plane. One of them tends
to a DeSitter attractor at infinity, defining a escape of orbits to the
inflationary regime. In fact, a straightforward analysis of the infinity of the
phase space shows the presence of a pair of critical points in this region,
one acting as an attractor (stable DeSitter configuration) and the other as
a repeller (unstable DeSitter configuration). The scale factor approaches
the DeSitter attractor as $a(\tau)\sim (C_0-\tau)^{-1}$ for $\tau \rightarrow C_0$,
where $\tau$ is the conformal time, or as $a(t) \sim {\rm exp} \Big(t \sqrt{\Lambda_4/3} \Big))$
for $t \rightarrow \infty$, where $t$ is the cosmic time.

\section{Preinflationary Model with Dark Matter, Dark Energy and Radiation}
To proceed with our analysis further we will consider braneworld models
whose matter content is restricted to a dark matter component (dust) and radiation,
together a massive scalar field (the inflaton) and a dark energy component described
by the effective cosmological constant in the brane (cf. Eq. (\ref{eq11rri})).
The reasons for these restrictions are two-fold. First the model contains a minimal set
of ingredients that is appropriate for a preinflationary model. The second is more
technical and has to do with the number of independent perfect fluid components.
A large number of fluid components results in a higher dimensional parameter space,
what turns the numerical/analytical analysis of parametric resonance quite
involved. Furthermore the resonance patterns that are important for the physics
of inflation in this model are typical for the dynamics of the general
model as we will discuss later. In this instance Eqs. (\ref{eq18})-(\ref{eq19}) reduce
then to
\begin{widetext}
\begin{eqnarray}
\label{eq27}
\nonumber
\frac{p_{a}^2}{12}+V(a)-|\sigma|E_{\rm rad}-\frac{|\sigma|}{2}\Big( {{\varphi}^{\prime}}^2+(1+m^2 {a}^2) {\varphi}^2\Big)
+\frac{1}{2a^2}\Big(\frac{E_{\rm rad}}{a^2}+\frac{E_{\rm dust}}{a} \Big)\Big( {{\varphi}^{\prime}}^2+(1+m^2 {a}^2) {\varphi}^2\Big)\\
+\frac{1}{8 a^4}\Big( {{\varphi}^{\prime}}^2+(1+m^2 {a}^2) {\varphi}^2\Big)^2=0,
\end{eqnarray}
\end{widetext}
and
\begin{eqnarray}
\label{eq28}
{\varphi}^{\prime\prime}+(1+m^2{a}^2)~ \varphi=0,
\end{eqnarray}
where
\begin{eqnarray}
\label{eq28r}
V(a)=3 a^2- \Lambda_{4} a^4-|\sigma|E_{\rm dust}a +\frac{1}{2} \Big(\frac{E_{\rm rad}}{a^2}+\frac{E_{\rm dust}}{a} \Big)^2.~~
\end{eqnarray}
We remark that for the integrable case $m=0$, Eq. (\ref{eq27}) simplifies to
\begin{eqnarray}
\label{eq27r}
\frac{p_{a}^2}{12}+{\tilde{V}}(a)-|\sigma| \Big(E_{\rm rad}+ {\mathcal{E}}_{\varphi}^{0}\Big)=0,
\end{eqnarray}
where
\begin{eqnarray}
\label{eq27rr}
\nonumber
{\tilde{V}}(a)&=&3 a^2-\Lambda_4 a^4 -|\sigma|E_{\rm dust}a \\
&+&\frac{1}{2}\Big(\frac{E_{\rm rad}+ {\mathcal{E}}_{\varphi}^{0}}{a^2}+\frac{E_{\rm dust}}{a} \Big)^2,
\end{eqnarray}
with ${\mathcal{E}}_{\varphi}^{0} \equiv ({{\varphi}^{\prime}}^2+{\varphi}^2)/2$ a constant of motion.
If we compare Eq. (\ref{eq27r}) with Eq. (\ref{eq23}) for the invariant plane,
and Eq. (\ref{eq27rr}) with Eq. (\ref{eq28r}) we may see that the integrable dynamics ($m=0$)
is analogous to the integrable dynamics in the invariant plane up to the substitution
$E_{\rm rad} \rightarrow E_{\rm rad}+ {\mathcal{E}}_{\varphi}^{0}$. In other words, the integrable scalar
field behaves as a radiation fluid in respect to the dynamics of the scale factor, a fact that will be used
in Section VII, for the case of pure scalar field cosmology. We should remark that
in the low energy limit the cosmological constant on the brane $\Lambda_{4}$ may be
interpreted as the effective vacuum energy of the inflaton field and $\varphi$
are the spatially homogeneous expectation values of the inflaton fluctuations
about its vacuum state. When these fluctuations are considered small or initially small, namely,
taken near the invariant plane, we may neglect ${\mathcal{E}}_{\varphi}^{0}$ and the integrable
dynamics $m=0$ is approximately that of the invariant plane. The scalar field fluctuations will then
actually have the role of just triggering the resonances in the perfect fluid cosmologies.
\begin{figure}
\begin{center}
\includegraphics[height=11cm,width=8.5cm]{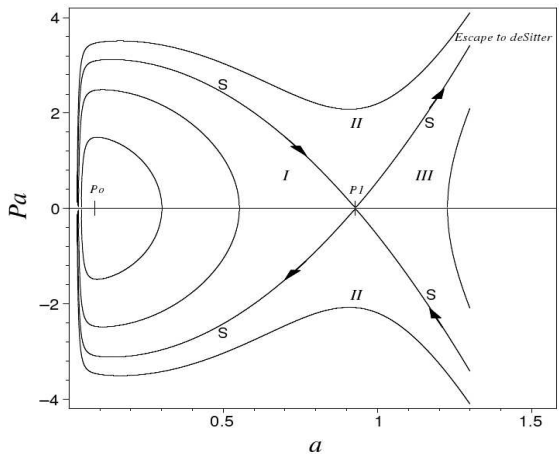}
\caption{The phase portrait of invariant plane dynamics with the critical points $P_0$ (center)
and $P_1$ (saddle-center) corresponding to stable and unstable Einstein universes.
The periodic orbits of region I describe perpetually bouncing universes.
Orbits in Region II are solutions of one-bounce universes. A separatrix $\mathcal{S}$ emerges
from $P_1$ towards the deSitter attractor defining a escape to inflation.}
\label{fig1}
\end{center}
\end{figure}
\par In Fig. 1 we depict the phase space portrait of the integrable dynamics in
the invariant plane ($\varphi=0=p_{\varphi}$) for the case of dust and radiation,
with $|\sigma|E_{\rm dust}$ sufficiently bounded so that $V(a)$ has a well
(we adopted $|\sigma|=500$ and $E_{\rm rad }=E_{\rm dust}=10^{-3}$).
The critical points $P_0$ (center)
and $P_1$ (saddle-center) correspond to stable and unstable Einstein universes.
Typically the model allows for the presence of perpetually bouncing universes
(periodic orbits) in region I of Fig. 1, associated with motion in the potential
well $V(a)$ about the stable
Einstein universe configuration $P_0$. These configurations are basically the
ones for which $|\sigma |~E_{\rm rad} < V(a_{\rm max})$. Region I,
understood as a nonlinear neighborhood of $P_0$, is bounded
by the homoclinic separatrix $\mathcal{S}$ emerging from the saddle-center $P_1$\cite{holmes}.
Orbits in Region II are correspond to one-bounce universes.
A separatrix emerges from $P_1$ towards the deSitter attractor at infinity, defining a
escape to inflation.
\par Finally it is worth mentioning here some basic structural differences
between the integrable dynamics in the invariant plane and that of the integrable case
(Eq. (\ref{eq27r}) and Eq. (\ref{eq28}) for $m=0$) whenever $\varphi(0)$ and/or $p_{\varphi}(0)$
are not zero. The phase space portrait in the plane ($a,p_a$) is similar to that of the invariant plane
shown in Fig. 1, however $P_1$ and $P_0$ are no longer critical points but periodic orbits.
The integrable dynamics is not restricted to this plane but, in the case of region I,
evolve on tori that are the direct product of the closed curves in region I to periodic
orbits in the sector ($\varphi, p_{\varphi}$). In particular the product of the separatrix
$\mathcal{S}$ that bounds the region I times a periodic orbit with initial conditions fixed by
the constant ${\mathcal{E}}_{\varphi}^{0}$ is a cylinder. The orbits on this cylinder coalesce
(for times $\tau \rightarrow \pm \infty$) to the periodic orbit at $P_1$ with the same constant of motion. This cylinder is said to be homoclinic to the periodic orbit at $P_1$.
\par As we mentioned already, for a fixed value of $\Lambda_4$ the potencial $V(a)$
has two extrema (one minimum and one maximum) for
suitably bounded values of ($|\sigma|,E_{\rm rad},E_{\rm dust}$),
corresponding to a well in the potential.
For the case of pure radiation we can show that $V(a)$ will have this well
provided that
\begin{eqnarray}
\label{eq27K}
E_{\rm rad}^{2} < \frac{2187}{2048}~\frac{1}{{\Lambda_4}^3}.
\end{eqnarray}
The two extrema of $V(a)$ for this case are critical points of the dynamics since
$V(a)>0$ for all $a>0$ implying that even the minimum of the potential belongs to the
physical phase space domain. For $E_{\rm rad}$ violating the above restriction no
extrema are present and the system has no critical point in the finite region of phase space.
For the case of pure dust $V(a)$ has no extrema if
\begin{eqnarray}
\nonumber
|\sigma|E_{\rm dust}>\frac{16 ~\sqrt{3}}{9}~\frac{1}{\sqrt{\Lambda_4}}.
\end{eqnarray}
For radiation plus dust we do not have a closed form for the constraint conditions
on the parameters. However it is not difficult to check that, in general, the increase
of $|\sigma|E_{\rm dust}$, namely, the increase of the
gravitational interaction strength of the dust component, has the effect of taking the
minimum of $V(a)$ out of the physical space leading eventually to a destruction
of the well. In other words the increase of $|\sigma|E_{\rm dust}$ has the effect of reducing
the phase space volume avaiable for bounded and/or initially bounded (metastable) configurations.
This is illustrated in Fig. 2 where we plot the potential $V(a)$
for several increasing values of $|\sigma|E_{\rm dust}$ with fixed $\Lambda_4$ and $E_{\rm rad}$
(as a matter of fact we fixed $\Lambda_4=1.5$, $E_{\rm dust}=0.001$, $E_{\rm rad}=0.01$ and varied $|\sigma|$).
We see that for $|\sigma|E_{\rm dust} \gtrsim 2.3$ the well in the potential is no longer present.
\begin{figure}
\begin{center}
\includegraphics*[height=11cm,width=8.5cm]{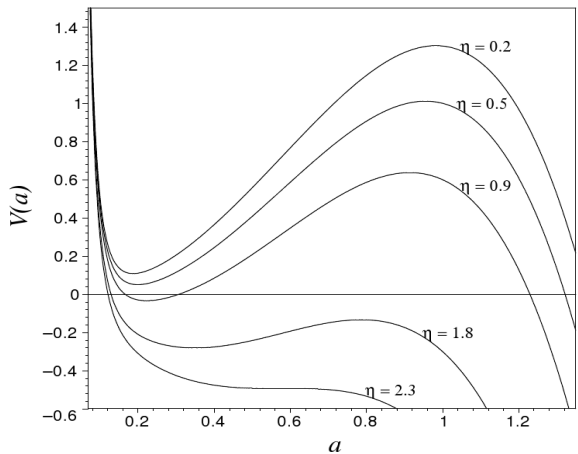}
\caption{Plots of the potential $V(a)$ for increasing values of $\eta \equiv |\sigma|E_{\rm dust}$.
For higher values of $|\sigma|E_{\rm dust}$ the minimum is out of the physical
phase space and eventually disappears (for $|\sigma|E_{\rm dust} \gtrsim 2.3$)
together with the potential well.}
\label{fig2}
\end{center}
\end{figure}
\par The structure of motion about $P_0$ will be
the main object of our interest in the following. This region is physically more
relevant than that of one bounce models since it avoids the theoretical problems of posing
initial conditions at past infinity, this issue being possibly brought to the realm
of a semi-classical quantization of the dynamics in the region I.
Furthermore maximum entropy considerations\cite{gibbons} favor the stable Einstein universe
as a suitable past configuration about which initial states of the
preinflationary universe oscillate.
However we must provide a mechanism for such bounded and
perpetually bouncing universe configurations to become metastable so that they may
realize inflation and escape to the DeSitter attractor at infinity. Such mechanism
will be the nonlinear resonance provided by the interaction with the scalar field sector
($\varphi,p_{\varphi}$) in the nonintegrable case, as we proceed to discuss.
\par We should finally remark that the increase of $|\sigma|E_{\rm dust}$, namely, the increase of
Newton's constant on the brane and/or the increase of the dust content of the model,
have the effect of taking the stable Einstein universe configuration out of the physical phase space
(cf. Fig. 2); in the latter situation the physical phase space is not simply connected
presenting a void with the structure of a solid torus.
\section{Nonlinear Resonance of KAM Tori}
\par We start by discussing the topology of motion about the stable Einstein universe $P_0$
that includes (i) the structure of KAM tori that are present in the neighborhood of $P_0$
and arise from the tori of the integrable case $m=0$; (ii) nonlinear resonance mechanisms
inducing that either KAM tori are destroyed by the resonance allowing the orbit escape to
the de Sitter attractor or the motion is stable (confined between two KAM tori) but
otherwise stable.
\par Let us consider the  energy surfaces
$|\sigma| (E_{\rm rad}+ {\mathcal{E}}_{\varphi}^{0}) < E_{\rm crit}(P_1)\equiv V(P_1)$
corresponding to bounded motion in the integrable case $m=0$, or to initially bounded motion
in the nonintegrable cases. This phase space region can indeed be characterized as a
nonlinear neighborhood of the center $P_0$, and is foliated by the two-tori
${\mathcal{S}}^1 \times {\mathcal{S}}^1$ that are the topological product of periodic
orbits of the separable sectors ($\varphi,p_{\varphi}$) and ($a,p_a$), with the two
associated separately conserved quantities ${\mathcal{E}}_{\varphi}^{0}$ and ${\mathcal{E}}_{a}^{0}=(p_{a}^{2}/12+{\tilde{V}}(a))$ satisfying ${\mathcal{E}}_{a}^{0}-|\sigma| (E_{\rm rad}+ {\mathcal{E}}_{\varphi}^{0})=0$. The frequency $\nu_a$ of the periodic orbit in the sector ($a,p_a$)
is given by the third-kind elliptic integral\cite{abramo}
{\small
\begin{eqnarray}
\label{eq29}
\frac{1}{\nu_a}=\sqrt{\frac{12}{\Lambda_4}}\int_{\beta_3}^{\beta_2}\frac{x^2~dx}{\sqrt{\Big(x^2-2\alpha_1x+(\alpha_1^2+\alpha_2^2)\Big)\prod_{i=1}^{6}(x-\beta_i)}},~~
\end{eqnarray}}
where ($\beta_i,i=1...6$) and ($\alpha_1 \pm i\alpha_2$) are, respectively, the real and imaginary roots of ${\tilde{V}}(a)-|\sigma| (E_{\rm rad}+ {\mathcal{E}}_{\varphi}^{0}) =0$ (~$\beta_3<\beta_2<\beta_1$ are the positive real roots). The two tori of the integrable case are the topological product of the above class of periodic orbits parametrized by ${\mathcal{E}}_{a}^{0}$ with the periodic orbits of the harmonic oscillator parametrized by ${\mathcal{E}}_{\varphi}^{0}$ (with frequency $\nu_{\varphi}=1/2\pi$).
\par For future reference we note that the periodic orbits of the sector ($a,p_a$) in the integrable
case will be represented by the elliptic fixed point ($\varphi=0,p_{\varphi}=0$), namely the origin
of the Poincar\'e map with surface of section $p_a=0$, in case ${\mathcal{E}}_{\varphi}^{0}=0$. For
${\mathcal{E}}_{\varphi}^{0} \neq 0$ the corresponding integrable tori are represented by closed invariant
curves about the origin of the map. For a small nonintegrable parameter $m$ this picture is
maintained with ($\varphi=0,p_{\varphi}=0$) as a center of a primary island of KAM tori; in fact
the KAM theorem\cite{kam} establishes the stability of tori with a sufficiently incomensurate frequency ratio, which in the present case means $\nu_a$ sufficiently
irrational. Other integrable tori are destroyed by the nonintegrable perturbation, and the region between
two remaining invariant tori presents an intricate dynamics (unstable periodic orbits, stable periodic
orbits surrounded by islands, broken separatrices, and stochastic layers, this structure repeating
down to smaller scales\cite{holmes}). The importance of KAM tori for Hamiltonian systems with
two degrees of freedom comes from the fact that they prevent the diffusion of trajectories in the
whole phase space, and thus preventing in our model the entrance to inflation. As $m$ increases numerical
experiments show that invariant KAM tori may be destroyed, with a consequent
loss of stability of the system. This is the case of interest to us as orbits initially trapped
about the center ($\varphi=0,p_{\varphi}=0$) can escape into the inflationary phase. An important
mechanism for this break-up of invariant tori is nonlinear resonance that occurs in a restricted
domain of parameters of the system as we proceed to examine using a semianalytical approach.
\par If $m$ is small and/or we start from an initial condition ($\varphi_0,p_{\varphi0}$) small
(namely, near the invariant plane) we may approximate Eq. (\ref{eq28}) as
\begin{eqnarray}
\label{eq30}
{\varphi}^{\prime\prime}+(1+m^2{a_0^2(\tau)})~ \varphi=0,
\end{eqnarray}
where $a_0(\tau)$ is a solution of ${{a}^{\prime \prime}}+dV(a)/da \approx 0$ (cf. Eq. (\ref{eq21})).
Eq. (\ref{eq30}) has the form of a Lam\'e-type equation, and therefore parametric resonance occurs
when the ratio
\begin{eqnarray}
\label{eq31}
R=\frac{\nu_a}{{\tilde{\nu}_{\varphi}}}
\end{eqnarray}
is a rational number, where\cite{correction}
\begin{eqnarray}
\label{eq31r}
{\tilde{\nu}}_{\varphi}=\frac{1}{2\pi}\sqrt{1+\frac{(0.9~m)^2}{2}~(\beta_{3}^{2}+\beta_{2}^{2})}.
\end{eqnarray}
Under this condition $\varphi$ begins to grow exponentially in time and to act on the
dynamics of the scale factor $a$ which in turn will modify (\ref{eq30}). This feedback
will restructure the resonance, either (i) leading the dynamics into a more unstable behaviour,
with amplification of the resonance mechanism and consequent breaking of the KAM tori
and escaping of the orbits; or (ii) leading the orbits to a general chaotic motion in a
bounded region of phase space. This general nonlinear resonance mechanism can be given an
approximate analytical treatment, through which we can fix the dominant resonances, namely,
the rational numbers $R$ corresponding to the dominant resonances. In this approximation,
where higher-order terms in ($\varphi,p_{\varphi}$) are neglected, the first integral (\ref{eq27})
is approximated by
\begin{eqnarray}
\label{eq32}
{\mathcal{H}} \equiv {\mathcal{E}}_{a}^{0}-|\sigma|{\mathcal{E}}_{\varphi}^{0}-\frac{|\sigma|}{2}{m^2 a^2 \varphi^2}\simeq |\sigma|E_{\rm rad}.
\end{eqnarray}
Further, in the remaining nonintegrable term the variable $a$ and $\varphi$ are substituted by the
integrable solutions $a_0(\tau)$ and $\varphi_0(\tau)$. Since the Hamiltonian character of the first integral is recovered in the approximation, we are led to introduce action-angle variables (${\cal{J}}_a, \Theta_a,{\cal{J}}_{\varphi},\Theta_{\varphi}$), the angle variables being defined as
($\Theta_a=\nu_a \tau,~\Theta_{\varphi}={\tilde{\nu}}_{\varphi} \tau$) such that both
$\Theta_a$ and $\Theta_{\varphi}$ vary in the interval $[0,1]$ during a complete cycle of the original variables.
We remark that in the numerical experiments considered we have ${\mathcal{E}}_{a}^{0} \approx |\sigma|E_{\rm rad}$ in the initial stages of the dynamics. ${\mathcal{E}}_{a}^{0}$ is however not conserved as the dynamics proceeds, the nonintegrable term being responsible for the exchange of energy with the sector
($\varphi,p_{\varphi}$). Taking into account that the function $a_0(\tau)$ is periodic with period $T_a=\nu_a^{-1}$, the expansion in circular functions of the nonintegrable term of (\ref{eq32}) takes the form\cite{abramo}
{\small
\begin{eqnarray}
\label{eq32r}
{\mathcal{H}}_i=-\frac{|\sigma|}{2}~m^2{\cal{J}}_{a}^{(0)}{\cal{J}}_{\varphi}^{(0)}\sum_{n} \Big(A_n \cos 2 n \pi \Theta_{a}\Big) \cos 4 \pi \Theta_{\varphi},~
\end{eqnarray}}
where $A_n$ are numerical coefficients depending the parameters of the model through the roots of $V(a)-|\sigma|E_{\rm rad} \sim 0$ (cf. Eq. (\ref{eq29})), and a zero superscript denotes action variables of the integrable case. Hamilton's equation for ${\cal{J}}_{a}$ derived from (\ref{eq32}) can be
integrated for each given term $n$ of the series, yielding to a first approximation
{\small
\begin{eqnarray}
\label{eq33}
\nonumber
{\cal{J}}_{a} \sim \frac{|\sigma|}{2}~m^2{\cal{J}}_{a}^{(0)}{\cal{J}}_{\varphi}^{(0)}\sum_{n} \frac{A_n}{2 \pi n {\tilde{\nu}}_{\varphi}}\Big[\frac{\cos (2 \pi n \Theta_{a}-4 \pi \Theta_{\varphi})}{R-2/n}\\
+\frac{\cos (2 \pi n \Theta_{a}+4 \pi \Theta_{\varphi})}{R+2/n} \Big].~
\end{eqnarray}}
From (\ref{eq33}) we have that the dominant terms are the ones for which $R \simeq 2/n$. Therefore
for a fixed $n \geq 2$ the expression
\begin{eqnarray}
\label{eq34}
R=\frac{\nu_a}{{\tilde{\nu}_{\varphi}}}=\frac{2}{n},~~~n \geq 2
\end{eqnarray}
determines a volume in the parameter space ($|\sigma|,E_{\rm rad},E_{\rm dust},m$)
in the neighborhood of which a $n$-resonance occurs. It
represents a further step beyond the analysis of resonances (\ref{eq31}) in the linear regime
of Lam\'e-type equation (\ref{eq30}). The set up of the resonance is signaled by
the bifurcation of the periodic
orbit at the origin ($\varphi=0,p_{\varphi}=0$) into an unstable periodic orbit plus one or two
characteristic stable periodic orbits of the resonance (according respectively whether
$n$ is odd or even), a fact crucial to realize inflation, as we will discuss later.
\par The resonance chart for the model can now be constructed numerically using the exact
dynamics. Expression (\ref{eq34}) -- where approximations as well as the neglect of non-resonant
terms were used -- constitutes an accurate guide to localize and label the resonances in
the parameter space (obviously for a fixed $\Lambda_4$, the values of $|\sigma|E_{\rm dust}$ and $E_0$
must be compatible with initially bounded motion).
However, in the actual chart of resonance constructed numerically using the exact dynamics,
these domains will have a spread that is a correction of the approximations due the full dynamics.
\par The focus on the underlying bulk-brane structure of the gravitational dynamics will
lead us to examine initially the plane ($\sigma,m$) of the parameter space, while fixing
the total mass-energy of dust $E_{\rm dust}$ and of radiation $E_{\rm rad}$
in several distinct ratios. For numerical purposes in all numerical experiments we
will fix $\Lambda_4=1.5$.
\par Fig. 3 displays the resonant chart in the plane ($\sigma,m$) of the
parameter space for fixed $E_{\rm dust}=E_{\rm rad}=10^{-3}$. The chart correspond to initial
conditions taken near the invariant plane, namely near the periodic orbit at the origin
($\varphi=0,p_{\varphi}=0$), with $p_a=p_{\varphi}=0,\varphi=10^{-4}$. The continuous lines
are solutions of (\ref{eq34}) while the gray regions spreading about the lines are sections
of the resonance windows by the planes $E_{\rm dust}=0.001,E_{\rm rad}=0.001$. The remaining (white)
regions are domains of stable motion, with the dynamics bounded by KAM tori. We restricted ourselves
up to the resonance $n=5$ in order to not overcrowd the Figure.
\begin{figure}
\begin{center}
\includegraphics*[height=11cm,width=8.5cm]{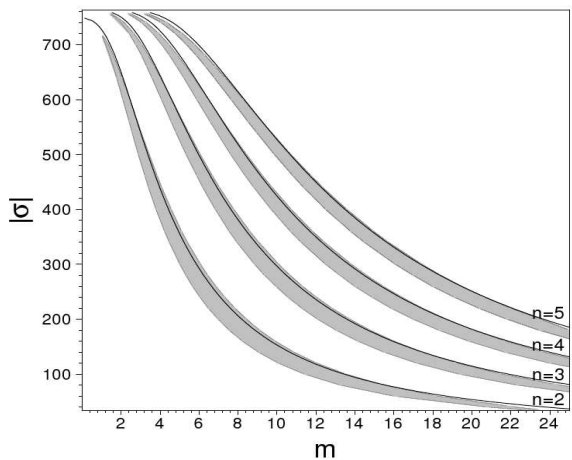}
\caption{Resonance chart in the plane ($\sigma,m$) for $E_{\rm rad}=E_{\rm dust}=10^{-3}$ and $\varphi_0=10^{-4}$.
The continuous lines are solutions of the approximate resonant condition (\ref{eq34}), while
the gray regions about the lines are the parametric resonance windows for the exact dynamics,
corresponding to the bifurcation of the periodic orbit at the origin. The white regions correspond
to KAM stable motion.}
\label{fig3}
\end{center}
\end{figure}
On driving the system towards a resonance zone (by an appropriate change of $m$ and/or $|\sigma|$)
we can turn a stable configuration into
a metastable with consequent escape of the orbits to the de Sitter attractor at infinity in a finite time.
In the realm of our pre-inflationary models, stability versus non-linear resonance
instability will be considered connected to initial conditions near the invariant plane only,
namely with ($\varphi,p_{\varphi}$) small, corresponding to spatially homogeneous fluctuations of
the inflaton field. Non-linear resonance will turn orbits generated from these initial conditions
from stable to unstable (and vice versa) as a consequence of bifurcation of the critical point
($\varphi=0,p_{\varphi}=0$),
at the origin of the Poincar´e map with surface of section $p_a = 0$, from a center to a saddle (and
vice versa), breaking up the KAM tori that trap the orbits about the origin. We recall that
the origin of the map is a periodic orbit of period $1/\nu_a$ in the ($a,p_a$) sector. This is
illustrated by the two Poincar\'e maps of Fig. 4 with surface of section $p_a=0$, for
$E_0=E_1=10^{-3}$ and $|\sigma|=500$ fixed, and $m=5.85$ in the domain of parametric
resonance $n=3$ (top map), and $m=7.1$ in the domain of parametric stability between
the resonances $n=3$ and $n=4$ (bottom map).
\begin{figure}
\begin{center}
\includegraphics[width=8.5cm,height=11cm]{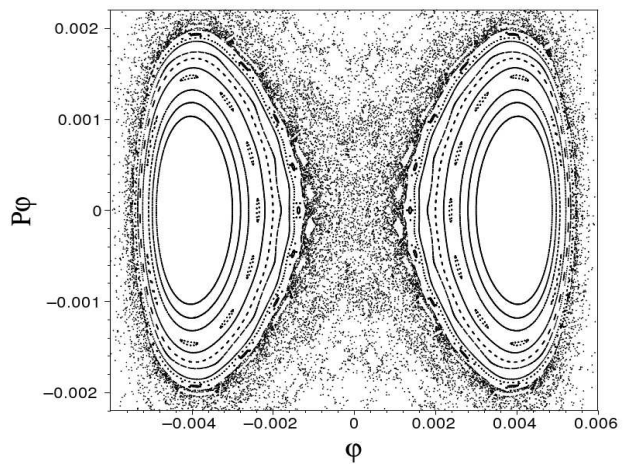}
\includegraphics[width=8.5cm,height=11cm]{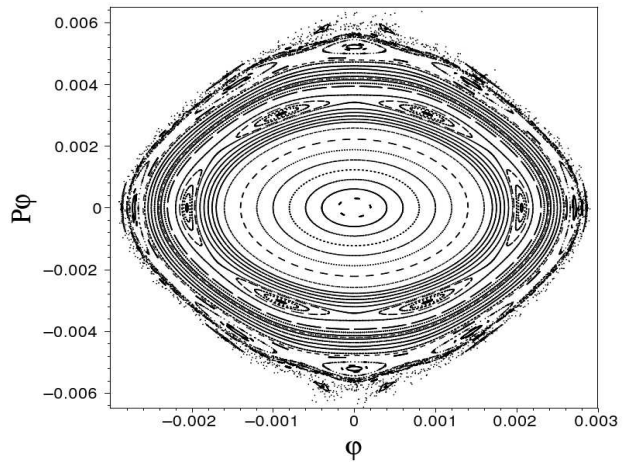}
\caption{Poincar\'e maps with surface of section $p_a=0$ for ($|\sigma|=500,m=5.85$) in the domain
of parametric resonance $n=3$ (top), and ($|\sigma|=500,m=7.1$) in the domain of parametric stability
between the resonances $n=3$ and $n=4$ (bottom). The origin in the top map is a saddle, connected to
the bifurcation of the origin due to the resonance $n=3$, favoring inflation. The bottom map has a center at the origin, enclosed by KAM tori that trap the inflaton, preventing inflation.}
\label{fig4}
\end{center}
\end{figure}
The origin in the top map is a saddle, with two associated centers, a consequence of the bifurcation
of the periodic orbit at the origin due to the resonance $n=3$. No KAM tori is present about the origin of the map so that orbits with initial conditions about the origin are not trapped and free to escape for large regions of phase space and eventually reach the deSitter attractor at infinity, realizing inflation.
The parametric domain of resonance thus favor inflation. The origin of the bottom map is a center,
enclosed by KAM tori that trap the orbits with initial conditions in this neighborhood, forbidding escape to the deSitter infinity. Therefore the region of parametric stability of the system will be unfavorable to the physics of inflation since the orbit (a configuration of the early universe) will be trapped in a stable state between two KAM tori about the center at the origin. The structure of the stochastic sea is distinct in each case. If the system is in the region of parametric resonance, initial conditions near the invariant
plane may undergo a long time diffusion through the stochastic sea to large regions of phase space, and
finally escape to the deSitter attractor. On the other hand, when the system is in the region of parametric
stability initial conditions for orbits that diffuse are far from the origin, beyond the borders of the
main island of the map, and diffusion with escape to deSitter infinity is extremely rapid.
\par Numerical experiments with the exact dynamics show that the presence of dust has the effect of
squeezing the volume of the resonance
windows, being thus less favorable to the occurence of inflation.
In Figure 5 we show the resonance windows $n=3$, corresponding to $\varphi_0=10^{-4}$, for
fluid content with pure dust ($E_{\rm rad}=0,E_{\rm dust}=0.001$), pure radiation
($E_{\rm rad}=0.001,E_{\rm dust}=0$) and a mixture of both ($E_{\rm rad}=E_{\rm dust}=0.001$),
the latter already shown in Fig. 3. Their range in the parameter space and relative size
are distinct, the avaiable width for a fixed $\sigma$ being much reduced when dust is present.
\begin{figure}
\begin{center}
\includegraphics*[height=11cm,width=8.5cm]{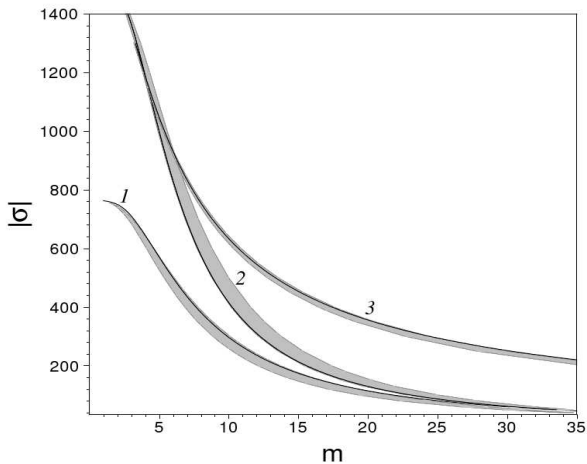}
\caption{Resonance windows $n=3$ for (1) dust plus radiation ($E_{\rm rad}=E_{\rm dust}=10^{-3}$),
(2) pure radiation ($E_{\rm rad}=10^{-3}, E_{\rm dust}=0$) and (3) pure dust
($E_{\rm rad}=0,E_{\rm dust}=10^{-3}$). The range and relative size of the windows are
distinct, the case of pure dust having a relative smaller volume though a more extended
region of parameters. The presence of dust has the effect of squeezing the width of the
resonance windows, reducing the domain of favorable configurations to realize inflation
as concerning the parametric resonance mechanism.}
\label{fig5}
\end{center}
\end{figure}
For reference we give the width $\Delta m$ of $n=3$ resonances windows for $|\sigma|=600$
and $|\sigma|$=400, and for several distinct matter content:
{\footnotesize{
\begin{center}
\begin{tabular}{|c|c|c|c|c|}
\hline
&  & & &\\
  & $E_{\rm rad}=10^{-3}$ & $E_{\rm rad}=10^{-3}$&$E_{\rm rad}=10^{-3}$ & $E_{\rm rad}=0$\\
& $E_{\rm dust}=0$& $E_{\rm dust}=10^{-4}$& $E_{\rm dust}=10^{-3}$&$E_{\rm dust}=10^{-3}$\\
 & & & &   \\\hline
$\Delta m$&1.1088& 1.0446& 0.51665& 0.55880 \\
{\tiny {$|\sigma|=600$}}& & & &  \\\hline
$\Delta m$&1.4418& 1.3850& 0.9307& 1.1305\\
{\tiny {$|\sigma|=400$}}& & & & \\\hline
\end{tabular}
\end{center}
}}
Hence if we demand that dark matter is present as a dust fluid in this preinflationary
phase then its total mass-energy content must not be exceedingly large in comparison with the radiation
content in order that inflation be properly realized. In addition we recall that, as discussed in Section IV,
the presence of dust has the effect of reducing the avaiable phase space volume for bounded
and/or initially bounded (metastable) configurations about the Einstein stable universe,
as regulated by $|\sigma|E_{\rm dust}$.
\par Summarizing,the instability versus the stability of the
origin ($\varphi=0,p_{\varphi}=0$) is crucial for the
dynamics of inflation, having a bearing on the dynamics
of the spatially homogeneous expectation values $\varphi(\tau)$ of the inflaton field
related to the escape into inflation. In this instance the initial conditions for $\varphi$
are assumed to be small, and are to be taken near the invariant plane (\ref{eq22}),
which corresponds to a neighborhood of the critical point of the map at the origin
($\varphi=0,p_{\varphi}=0$). Therefore the region of parametric stability is unfavorable
for producing inflation since the orbit (a configuration of the early universe) will
be trapped in a stable state enclosed by two
invariant tori of a main KAM island of the map. On the other hand, on driving the system
to a region of parametric resonance, orbits with initial conditions near the invariant plane
are turn into metastable configurations that either escape rapidly to de Sitter infinity or undergo
a long time diffusion through stochastic regions of phase space before finally escaping.
The resonance windows in the complete parameter space ($\sigma, E_{\rm rad}, E_{\rm dust},m$)
are constituted of $n$ disjoint 4-dim volumes the sections of which -- for instance with the
surfaces $E_{\rm rad}=10^{-3}$ and $E_{\rm dust}=10^{-3}$ -- result in the $n$ gray regions
of Fig. 3. The volumes of the windows are small as compared to
the whole volume of the parameter space, and only initial configurations inside them may
realize inflation. We note that for a fixed $m$ the larger the order $n$ of the resonance the stronger the
gravitational interaction in the braneworld universe inflated from
initial conditions connected with the resonance considered.
\section{A Partition in the resonace windows coded by disruptive resonances}

From the point of view of the dynamics of inflation, the resonance windows present a
further structure connected with disruptive resonances and/or long time diffusion before
escape to inflation. In fact, as we proceed to discuss, a considerable domain of the
resonance windows -- although corresponding to a bifurcation of the stable periodic orbit
at the origin -- does not lead to escape into inflation and must be properly discarded.
To simplify our analysis we restrict ourselves to the resonance window $n = 3$ (cf. Fig. 3)
at fixed $|\sigma|=350$. The associated range of $m$ lies in the interval
$\Delta m \cong [7.78431,8.84341]$. A careful numerical examination shows that
in this interval of the resonance zone we observe three dynamically distinct regions
are present:
(i) the domain on the left, ranging from the lower limit up to $m \lesssim 8.52$,
corresponds to configurations for which the dynamics is highly unstable,
the resonances being disruptive with a rapid escape to inflation (up to $\tau \leq 15,000$).
(ii) A threshold region $8.52< m < 8.56$ that corresponds to configurations of orbits
that undergo a long time diffusion ($15,000 < \tau < 100,000$) before escaping to inflation.
(iii)The region on the right of the threshold, ranging from $m \gtrsim 8.56$ up to the upper
limit of $\Delta m$. The motion of orbits connected to this latter parametric region --
although corresponding  to the case of a bifurcated saddle at the origin --
is resonant and chaotic, but otherwise stable. Therefore region (iii) should be discarded
since cosmological scenarios corresponding to these configurations do not
properly realize inflation.
\par The above dynamically distinct behaviors are illustrated as follows.
Fig. 6 shows the time signals for $m=7.99$ taken in the region of disruptive resonance;
the escape to inflation occurs at $\tau \simeq 140$ and time signals are used since there is
not enough recurrence for constructing a well defined Poincar\'e map.
Fig. 7 displays the Poincar\'e map with surface of section $p_a=0$ of a single orbit
for $m=8.555$, corresponding to the threshold region (ii), undergoing a long time
diffusion before exit to inflation, at $\tau \simeq 64,000$.
Finally the time signals shown in Fig. 8 correspond to $m=8.6$ in region (iii)
near the right border of the resonance window $n=3$. The motion is
resonant and chaotic, but otherwise stable. All Figures were generated with
$\varphi^{\prime}=0=p_a$, $\varphi=10^{-4}$.
\begin{figure}
\vspace{0.0cm}
\begin{center}
\hspace{0.0cm}
\vspace{0.0cm}
\includegraphics*[height=11cm,width=8.5cm]{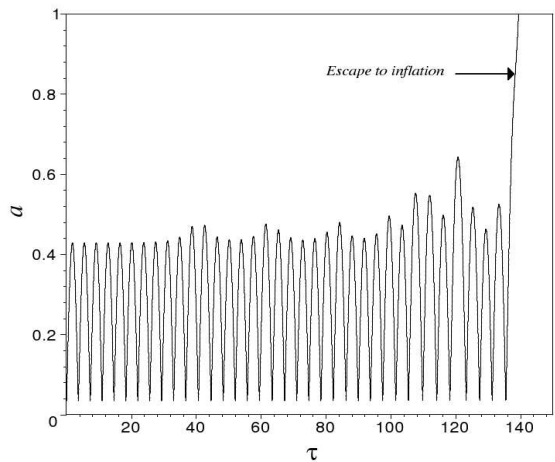}
\vspace{0.0cm}
\includegraphics*[height=11cm,width=8.5cm]{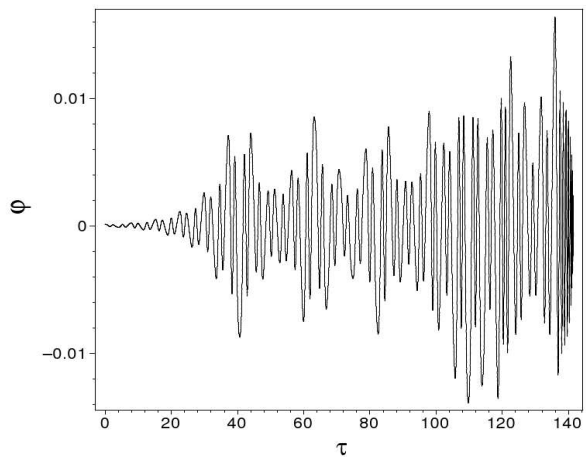}
\caption{Time signals for $m=7.99$ and $|\sigma|=350$ corresponding to the
region of disruptive resonances, near the left border of the $n=3$ resonance window
of Fig. 3.}
\label{fig6}
\end{center}
\end{figure}
\begin{figure}
\vspace{0.0cm}
\begin{center}
\hspace{0.0cm}
\vspace{0.0cm}
\includegraphics*[height=11cm,width=8.5cm]{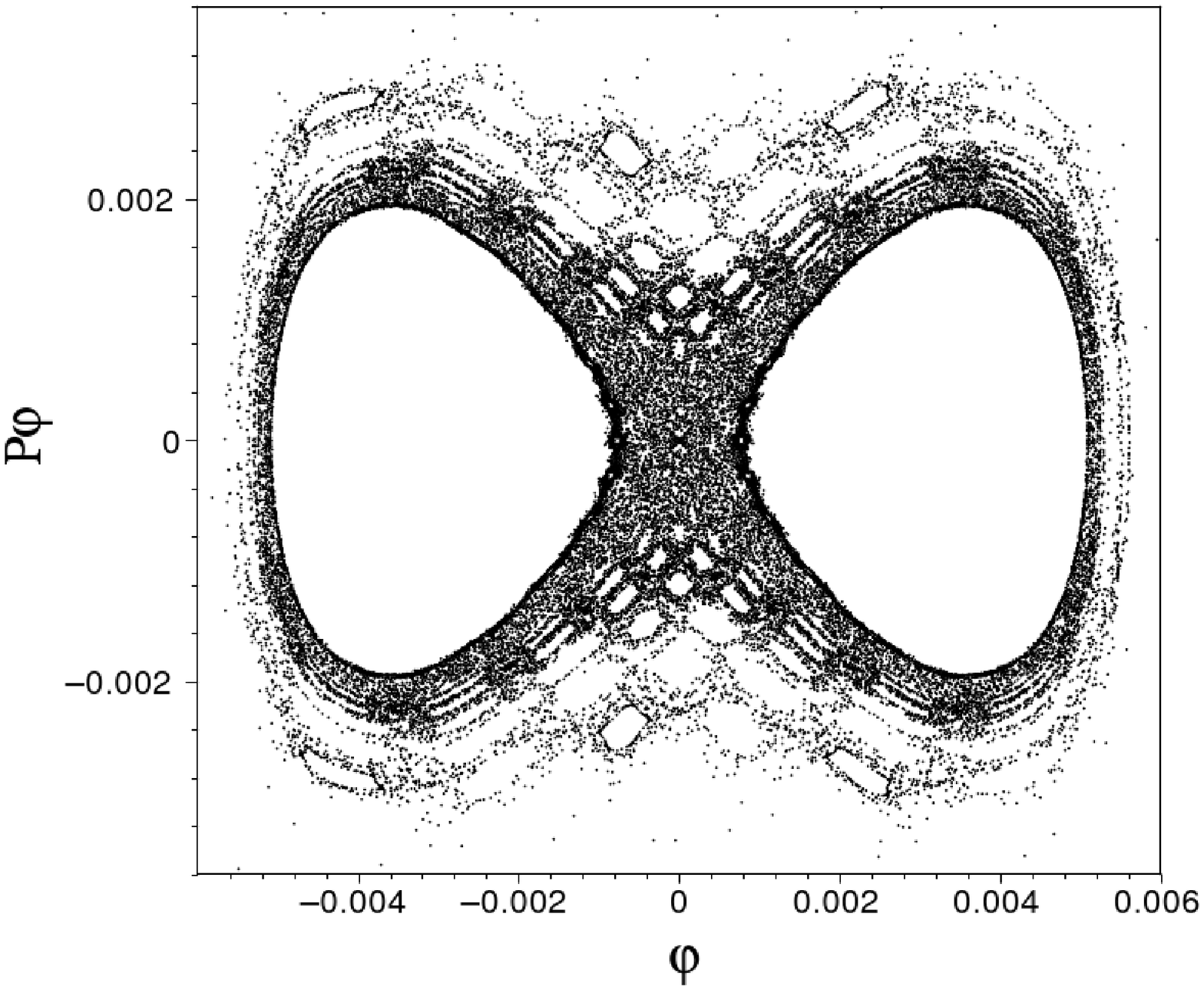}
\caption{Poincar\'e map with surface of section $a^{\prime}=0$ of
a single orbit, for $m=8.555$ and $|\sigma|=350$ in the threshold region of the $n=3$
resonance window of Fig. 3. The orbit undergoes a long time diffusion before escape
to inflation at $\tau \simeq 64,000$. The map exhibits the structure of the random motion
of the orbit in the stochastic sea about primary and secondary KAM islands of the resonance.}
\label{fig7}
\end{center}
\end{figure}
\begin{figure}
\vspace{0.0cm}
\begin{center}
\hspace{0.0cm}
\vspace{0.0cm}
\includegraphics*[height=11cm,width=8.5cm]{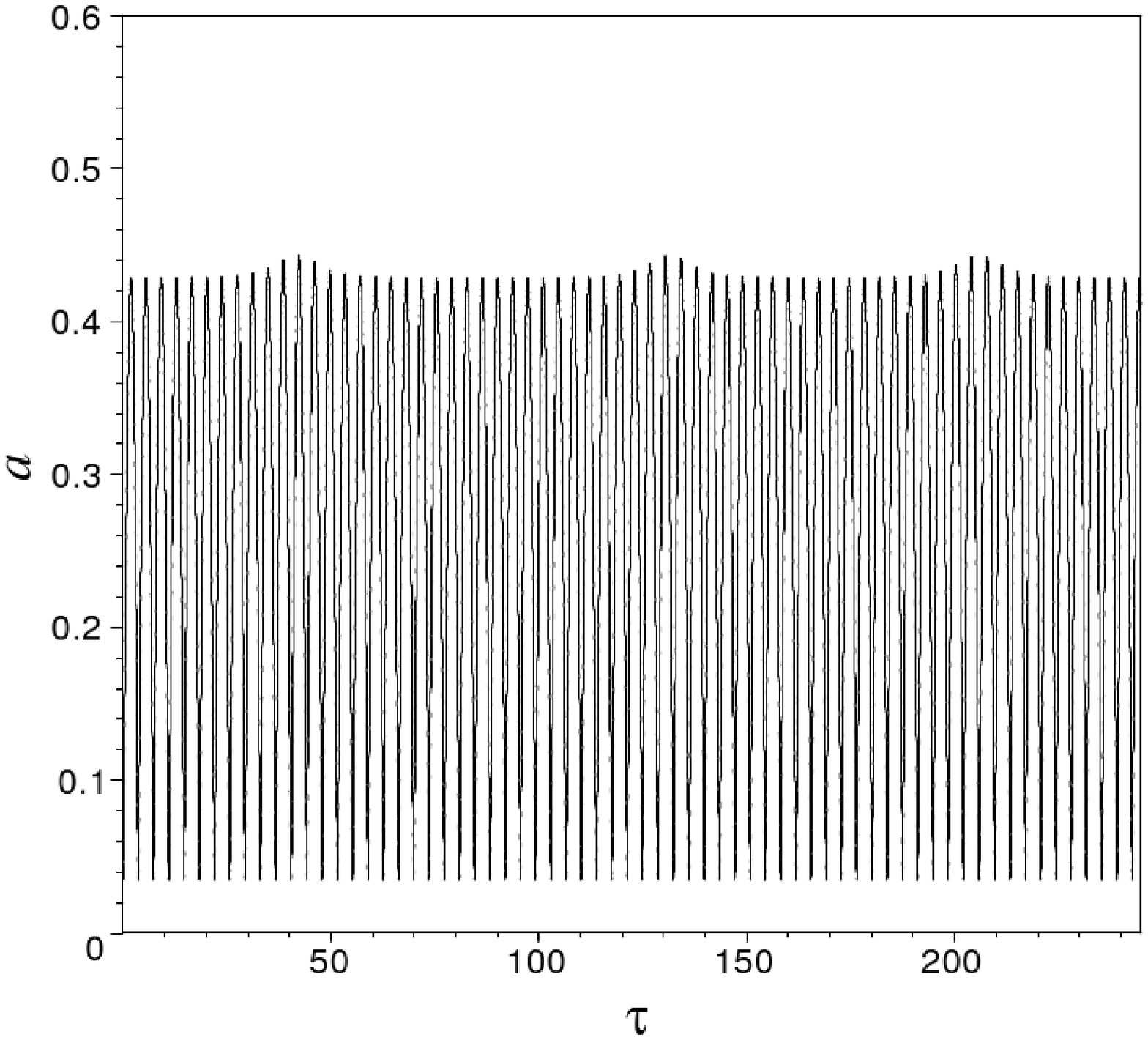}
\vspace{0.0cm}
\includegraphics*[height=11cm,width=8.5cm]{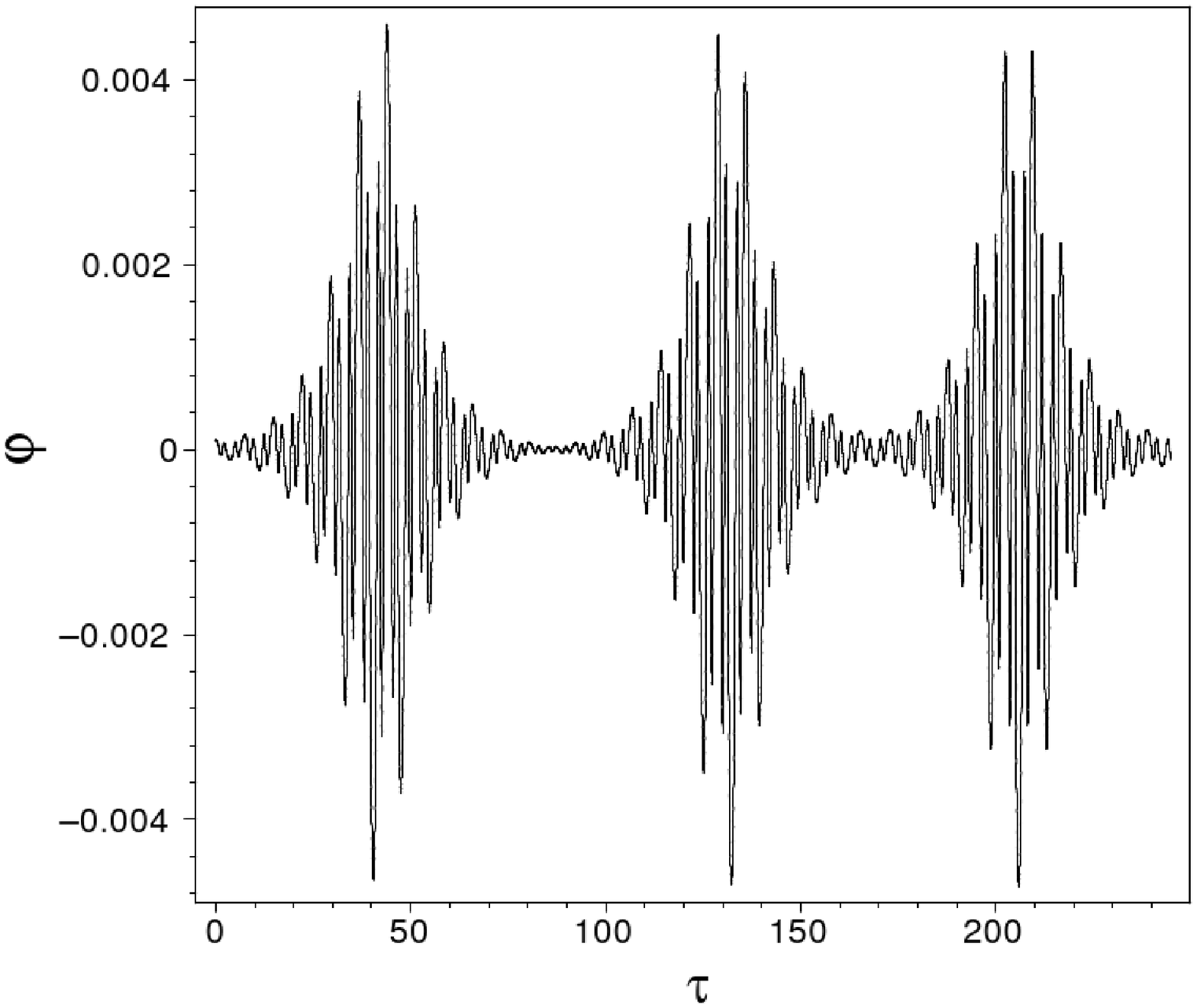}
\caption{Time signals for $m=8.6$ and $|\sigma|=350$, corresponding to the
region (iii) near the right border of the $n=3$ resonance window of Fig. 3.
The motion is resonant and chaotic, but otherwise stable, with no
exit to inflation.}
\label{fig8}
\end{center}
\end{figure}
\par The above substructure appears to be a feature of the whole resonance window, as we have checked
for other values of $|\sigma|$ in the $n=3$ window, as well as for other resonance windows.
It is worth mentioning that in the case of pure dust the subdomain of disruptive resonances
is a very thin sheet on the left border of the resonance window, while the larger portion of the
window that allows for escape corresponds to long time diffusion.
\par Finally we should remark that -- as concerning nonlinear resonance phenomena --
the general picture is that bouncing oscillating braneworld models
have a small restricted domain in their parameter space where inflation
can be realized. Typically variation of the parameters can shrink or stretch the resonance zones.
However the underlying pattern of resonance windows and their internal substructure is maintained
as we have checked numerically. In this sense the pattern is said to be structurally stable.
\section{Pure Scalar Field Bouncing Cosmologies. Metastable Dynamical Confinement}
We now proceed to examine the case when the model has no perfect fluid component.
As we will see, even then the bulk-brane corrections allow for the presence of
either perpetually oscillatory or metastable oscillatory models that emerge into an inflationary
phase after a finite time. The dynamical equations in this case are given by Eq. (\ref{eq19}),
namely the KG equation,
\begin{eqnarray}
\nonumber
{\varphi}^{\prime\prime}+(1+m^2{a}^2)~ \varphi=0,
\end{eqnarray}
plus the modified Eq. (\ref{eq18}),
\begin{eqnarray}
\label{eq35}
{{a}^{\prime}}^2+{a^2}-\frac{\Lambda_4}{3}~a^4+\frac{1}{6a^4} \Delta^{2}(\varphi) - \frac{|\sigma|}{3} \Delta(\varphi)=0,
\end{eqnarray}
where
\begin{eqnarray}
\label{eq36}
\Delta(\varphi)=\frac{1}{2}\Big[ {{\varphi}^{\prime}}^2+(1+m^2 {a}^2) {\varphi}^2\Big].
\end{eqnarray}
For the integrable case $m=0$, we have that $\Delta_0 \equiv [{{\varphi}^{\prime}}^2(\tau)+{\varphi}^2(\tau)]/2
\equiv [{{\varphi}^{\prime}}^2(0)+{\varphi}^2(0)]/2$
is a constant of motion and Eq. (\ref{eq35}) reduces then to
\begin{eqnarray}
\label{eq37}
{{a}^{\prime}}^2+V(a)=\frac{|\sigma|}{3} \Delta_0,~~~V(a)={a^2}-\frac{\Lambda_4}{3}~a^4+\frac{\Delta_0^{2}}{6a^4},~~
\end{eqnarray}
so that the dynamics of the scale factor $a(\tau)$ is analogous to that of a radiation dominated
bouncing braneworld universe with total energy content $\Delta_0$ (cf. Eqs. (\ref{eq18}) and (\ref{eq19})
for $E_{\rm rad}\neq 0$, other $E_i$=0). Physical motion imposes the constraint
$V(a_{\rm min}) \leq |\sigma|\Delta_0$, where $V(a_{\rm min})$ is the minimum of the potential (\ref{eq37}),
the inequality corresponding to oscillatory motion.
A distinct feature of the phase space is that the infinite barrier avoiding the singularity
as well as the constant of motion $|\sigma|\Delta_0$ are both built up with the
nonzero initial amplitudes of the inflaton field. A portrait of the phase space plane
($a^{\prime},a$) -- analogous to that in Fig. 2 -- may be constructed by varying $|\sigma|\Delta_0$.
In the domain $V(a_{\rm min}) \leq |\sigma|\Delta_0 < V(a_{\rm max})$ the dynamics lies on tori that are
the product of the periodic orbits in this phase plane times the circles
${{\varphi}^{\prime}}^2+{\varphi}^2=2 \Delta_0$. The minimum configuration is a
stable Einstein universe sourced by a massless inflaton.
This stable Einstein universe is a limiting 1-dim torus, being actually the product of
the point ($a^{\prime}=0, a=a_{\rm min}$) times the circle defined by
$\Delta_0 = V(a_{\rm min})/|\sigma|$. Contrary to the cases with a perfect fluid component,
the Einstein universe is not a critical point of the dynamics.
No invariant plane is present.
\par The compact phase space domain containing tori is analogously restricted by
(\ref{eq27K}) with the obvious substitution of $E_0$ by $\Delta_0$.
We note that, for fixed $\sigma$ and $\Lambda_4$, each of the integrable tori is
generated by just one curve with initial conditions (${\varphi}^{\prime}_0,\varphi_0$).
In this sense the energy surfaces $|\sigma|\Delta_0$=const. contain just one curve.
\par Analogous portrait can be constructed for a single orbit with initial conditions
(${{\varphi}^{\prime}}_0,{\varphi}_0$) by varying $\sigma$,
characterizing thus universes with distinct gravitational strengths $G_N=|\sigma|/8 \pi$.
The corresponding tori yield qualitatively the same stable dynamics, about the stable
Einstein universe sourced by a massless conformal inflaton with a minimal
gravitational strength.
\par For the nonintegrable case, with $m$ small (and
$|\sigma|m^2\varphi(0)^2/6$ sufficiently smaller than $1$, as we discuss below),
the pattern is that of stable motion on KAM invariant tori or between two invariant KAM
tori resulting from the integrable case. The stability of motion implies that all orbits
are trapped between invariant tori and cannot escape to inflation, namely,
inflation cannot be realized for these configurations. This is illustrated in Fig. 9 where
we construct the Poincar\'e map with surface of section $a^{\prime}=0$
for fixed parameters ($\Lambda_4=1.5, |\sigma|=500, m=8.15$).
\begin{figure}
\vspace{0.0cm}
\begin{center}
\hspace{0.0cm}
\vspace{0.0cm}
\includegraphics*[height=11cm,width=8.5cm]{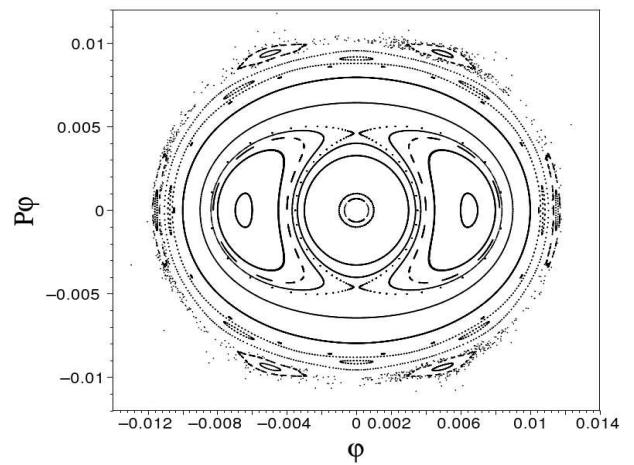}
\caption{Poincar\'e map with surface of section $a^{\prime}=0$ for pure
scalar filed cosmologies, with $|\sigma|=500$ and $m=8.15$. The innermost
circle about the origin is a 1-dim torus corresponding to a stable Einstein universe sourced by a scalar field.
The domain inside this circle is unphysical. The last KAM torus shown has initial conditions
($\varphi^{\prime}_0=0, \varphi_0 \simeq 0.01105$). Escape to inflation may occur in the outer border
(for $\varphi_0 > 0.01105$) with oscillatory motion due to partial confinement of orbits.}
\label{fig9}
\end{center}
\end{figure}
The innermost circle about the origin of the map is the section of the 1-dim torus
(namely, the periodic orbit in the plane (${\varphi}^{\prime},\varphi$)~)
corresponding to the stable Einstein universe. In this example the stable Einstein
universe -- sourced by a massive ($m=8.15$) conformal inflaton -- is described by
$a^{\prime}=0,a \simeq 0.00519690052$) times the circle in the
(${\varphi}^{\prime},\varphi$)-plane generated, for instance, from the initial conditions
(${\varphi}^{\prime}(0)=0$, $\varphi \simeq 0.000696503226$) with associated energy constant
$\Delta_0 \simeq 0.2429935 \times 10^{-6}$. Below this scalar field energy no oscillatory motion
is found, resulting in the void (no physical motion) about the origin of the map
bounded by the above described innermost circle.
The pattern of KAM tori sections, with eventual bifurcated secondary islands,
extends up to initial conditions $a^{\prime}=0$, $\varphi^{\prime}=0$ and $\varphi_0 \simeq 0.01105$
where approximately the last confining
KAM torus lies. In this region of trapped orbits we may note, for instance,
$8$ secondary islands (the far right centered about $\varphi_0=0.01079$)
connected to a $8/5$ bifurcation. The outside region, denoted the stochastic
sea in border of the main KAM island, corresponds to
the domain of initial conditions where inflation can be realized in pure
scalar field sourced braneworld cosmologies. In this stochastic sea we may have either
a rapid escape to inflation (for $\varphi_0=0.0118$ with $\tau \simeq 210$, or
for $\varphi_0=0.01125$ with $\tau \simeq 500$, for instance) or a diffusion with escape,
for instance about the border of the two sets of $3$ secondary islands connected
to a $3/2$ bifurcation (for $\varphi_0= \pm 0.011305$ with $\tau \simeq 1690$).
For larger initial conditions, namely
$\varphi_0 \gtrsim 0.01234$, no bifurcated islands are found in the stochastic sea
but finite time oscillations are still found due to a mechanism of dynamical partial
confinement, connected to values of the parameter
\begin{eqnarray}
\label{eq38}
\varsigma \equiv |\sigma|m^2 \varphi(0)^2/6
\end{eqnarray}
sufficiently close to $1$, as we describe in the following.
\par Let us recall that the potential $V(a)$ in (\ref{eq37}) presents a well
(with possible oscillatory motion) for properly bounded values of $\Delta_0$ restricted by (\ref{eq27K}).
Furthermore $V(a)-|\sigma|\Delta_0/3=0$ must have three real positive
roots such that oscillatory motion is present. However the restriction (\ref{eq27K}) was derived for the
integrable case and for the spatial curvature parameter rescaled to $k=1$.
Actually the presence of the physical well is due to the balance between the
spatial curvature term and the infinite barrier term originated from the bulk-brane
correction, more specifically, the balance between the values of the parameters $k$ and $\Delta_0$.
It is not difficult to see that, in the integrable case, the decrease (increase) of $k$ with
$\Delta_0$ fixed as well as the increase (decrease) of $\Delta_0$ for $k$ fixed, may destroy
(or create) a well. Now for the nonintegrable case ($m \neq 0$) a careful examination of
the integral constraint (\ref{eq35}) shows that the term proportional to $m^2$ in the RHS will
contribute to correct the curvature term to an effective spatial curvature ($k_{\rm eff} \simeq 1-\varsigma$)
at the initial times. For the parameter configuration of Fig. 9 numerical evaluations show that
the above effects will be crucial for the dynamics of the scale factor $a(\tau)$ when $k_{\rm eff}\lesssim 0.157$, that corresponds to $\varphi_0 \gtrsim 0.01234$. In this instance, two possibilities arise: (i) for
$0.01234 \lesssim \varphi_0 < 0.1334$, $\varphi_0^{\prime}=0$, the function
\begin{eqnarray}
{\mathcal{P}}(a) \equiv {a^2}-\frac{\Lambda_4}{3}~a^4+\frac{1}{6a^4} \Delta^{2}(\varphi_0) -\frac{|\sigma|}{3} \Delta(\varphi_0),
\label{eq38}
\end{eqnarray}
(cf. (\ref{eq35})) has two extrema (one maximum and one minimum) but ${\mathcal{P}}(a)=0$ has only one
real positive root $a_0$, meaning that the value of the relative maximum of ${\mathcal{P}}(a)$ is smaller
than zero. The orbit evolved from this initial condition would in principle be expected to escape
without bounce. However, due to the increase of the effective time-dependent spatial curvature
$k_{\rm eff} \simeq 1-|\sigma|m^2 \varphi(\tau)^2/6$,
the relative maximum of ${\mathcal{P}}(a)$ is raised above zero at a later time so that the orbit
bounces back. This is illustrated in Fig. 10(a) where we plot ${\mathcal{P}}(a)$ for the initial
time with (${\varphi^{\prime}}_0=0,\varphi_0=0.0126$) which has the only real root
$a_0 \simeq 0.016955927198$ (continuous curve), and for the time of the first bounce of the
orbit generated from these initial conditions (dashed  curve). The first bounce corresponds to the point
(${\varphi^{\prime}}_0 \simeq -0.0144932,\varphi_0\simeq -0.00419929$)) with
$k_{\rm eff} \sim 0.9$. We note that the dashed curve represents a
dynamical potential with its maximum above zero, allowing for the bounce.
The adjustment of this effect with the period of the massive scalar field allows for
a series of bounces of the orbit before it finally escapes, as shown in Fig. 10(b). (ii) For
$\varphi_0 \gtrsim 0.1334$ ${\mathcal{P}}(a)$ has no extremum but due to the effective time-dependent
curvature two extrema are dynamically created, with the relative maximum above zero, so that the
orbit undergoes bounces in a relatively smaller number than in case (i) before escaping.
It is worth mentioning that this effect may be present for $\varphi_0 < 0.01234$, provided
we properly increase the parameters $m$ and/or $|\sigma|$ to turn $\varsigma$ sufficiently small.
\begin{figure}
\vspace{0.0cm}
\begin{center}
\hspace{0.0cm}
\vspace{0.0cm}
\includegraphics*[height=11cm,width=8.5cm]{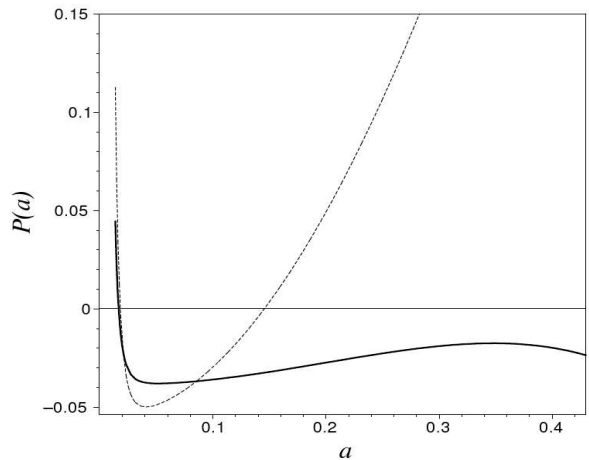}
\vspace{0.0cm}
\includegraphics*[width=8.5cm,height=11cm]{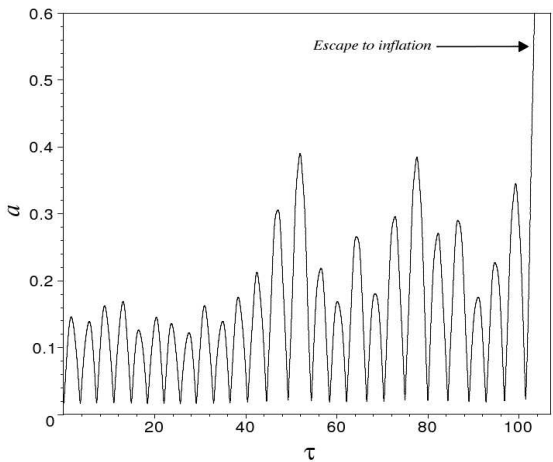}
\caption{Plot of the function ${\mathcal{P}}(a)$ for the initial time (continuous curve) and the time of the
first bounce (dashed curve). The occurrence of the bounce is due to the dynamical raising of the potential
as a consequence of the increase of the effective spatial curvature $k_{\rm eff} (\tau)$ as $\varphi$ decreases.
The first bounce occurs when $k_{\rm eff} \sim 0.9$. A sequence of this process
produces the oscillatory behaviour shown below.}
\label{fig10}
\end{center}
\end{figure}
\par Finally we should remark that the dynamics of pure scalar field brane cosmologies
is not connected to a parametric resonance pattern. The features of initial condition domains
that realize inflation in this case are similar to the ones for the domain of parametric
stability (compare the Poincar\'e maps  in Fig. 4(bottom) and Fig. 9), as the configurations
that realize inflation are the ones at the border of the primary KAM islands, corresponding
to large values of $\varphi$.
\section{Motion about The Saddle-Center. Homoclinic Chaos and the Chaotic Exit to Inflation}
The saddle-center critical $P_1$ induces, in the phase space of the models, the topology
of stable and unstable homoclinic cylinders which emanate from unstable periodic orbits
that exist in their neighborhood. This structure was examined in some detail in \cite{monerat}
and is briefly discussed here for completeness. The nonintegrability of the dynamics induces the
breaking and crossing of the homoclinic cylinders leading to a chaotic exit to inflation as we
proceed to show in this Section, where for simplicity we restrict ourselves to two perfect fluid
components (dust and radiation only) as in Section IV.
\par Our starting point is a fudamental property of saddle-centers given by
Moser's theorem\cite{moser1}. Moser's result states that it is always possible to find a set
of canonical variables such that in a small neighborhood of a saddle-center the energy
integral of motion (\ref{eq18}) is separable into rotational and hyperbolic motion pieces.
The variables ($a,p_a,\varphi,p_{\varphi}$) of our system are already of the Moser type
and, in a linear neighborhood of $P_1$, we have the separable rotational motion energy piece
$E_{(\varphi)}$ and the hyperbolic motion energy piece $E_{(a)}$ as expressed in (\ref{eq26}).
We recall that in this case $V^{\prime \prime}(a_{\rm crit}) < 0$.
In this approximation, we note that the scale factor $a(\tau)$ has pure hyperbolic motion and is
completely decoupled from the pure rotational motion of the inflaton fluctuation $\varphi$.
Let us consider the possible motions in this neighborhood. In the case
$E_{a}=0$ and $p_a=0=(a-a_{\rm crit})$, the motion corresponds to unstable periodic orbits
${\mathcal{T}}_{|\sigma|E_{\rm rad}}$ in the plane ($\varphi,p_{\varphi}$). Such orbits
depend continuously on the parameter $|\sigma|E_{\rm rad}$. For $E_{a}=0$
there is still the possibility $p_a= \pm \sqrt{6|V^{\prime \prime}(a_{\rm crit})|}~(a-a_{\rm crit})$,
which defines the linear stable $V_S$ and unstable $V_U$ 1-dim manifolds. The direct product
of the periodic orbit ${\mathcal{T}}_{|\sigma|E_{\rm rad}}$ with $V_S$ and $V_U$ generates, in the
linear neighborhood of the saddle-center, the structure of pairs of stable
(${\mathcal{T}}_{|\sigma|E_{\rm rad}} \times V_S$) and unstable
(${\mathcal{T}}_{|\sigma|E_{\rm rad}} \times V_U$) 2-dim cylinders. We note that for
$E_{\rm crit} \approx |\sigma|E_{\rm rad}$ the manifolds $V_S$ and $V_U$ are tangent at $P_1$
to the separatrices $\cal{S}$ of the invariant plane containing the saddle-center; hence
a pair of cylinders (one stable and one unstable) emanates from the neighborhood of $P_1$ towards
$a \sim 0$, while another pair emanates towards the two deSitter attractors at infinity (cf. Fig. 1).
Orbits on the cylinders coalesce into the periodic orbit
${\mathcal{T}}_{|\sigma|E_{\rm rad}}$ asymptotically, the orbits being contained
in the same energy surface $|\sigma|E_{\rm rad}$ as that of the periodic orbit; these orbits are denoted homoclinic
to the periodic orbit ${\mathcal{T}}_{|\sigma|E_{\rm rad}}$ or simply homoclinic orbits.
Noticing that in general $E_{(a)}-|\sigma|E_{(\varphi)}+ E_{\rm crit}-|\sigma|E_{rad} \sim 0$
(cf. (\ref{eq26s}) and (\ref{eq26})) and that $E_{\varphi}$ is strictly positive we must
have $E_{\rm crit} - |\sigma|E_{\rm rad} >0$;
hence only energy surfaces satisfying this condition contain homoclinic cylinders.
\par The nonlinear extension of the plane of rotational motion (when ${\mathcal{O}}(3)$ terms in (\ref{eq26s})
are taken into account) is a 2-dim manifold, the {\it center manifold} of unstable periodic orbits of
the system parametrized with the energy $|\sigma|E_{\rm rad}$. The intersection of the center manifold
with the energy surface $|\sigma|E_{\rm rad}$ is a periodic orbit ${\mathcal{T}}_{|\sigma|E_{\rm rad}}$
from which two pairs of cylinders emanate, as in the linear case previously described. It can be shown that in general the nonlinear extension of the center manifold folds, so that the unstable periodic orbits are no longer contained in the plane ($\varphi,p_{\varphi}$). Now the extension of the cylinders away from the periodic orbits depends crucially on the integrability or nonintegrability of the system.
\par In the integrable case ($m=0$) the homoclinic cylinders are the topological product of the
separatrices $\mathcal{S}$ times the periodic orbit with energy ${\mathcal{E}}^{0}_{\varphi}$.
We are basically interested in the pair of cylinders (one stable and one unstable) that
emerge from the periodic orbit towards decreasing values of $a$, namely, in region I.
It is not difficult to see that, in the integrable case, the unstable cylinder coalesces smoothly into the stable one. However when the nonintegrability is switched on (namely, $m \neq 0$), the smooth continuation
of the unstable cylinder into the stable one breaks, inducing the transversal crossing of them. The points
of intersection define homoclinic orbits which emerge from the periodic orbit along the unstable
cylinder and return to it along the stable one, in an infinite time. Typically if the 2-dim cylinders
intersect transversally once, they will intersect each other an infinite number of times, producing
an infinite set of homoclinic orbits that are bi-asymptotic to the unstable periodic orbit of the
center manifold. This infinite set of homoclinic orbits is denoted the intersection manifold.
The dynamics near homoclinic orbits is very complex, associated with the presence of the well-known
horseshoe structures (cf. \cite{holmes}, \cite{wiggins} and references therein), with the homoclinic intersection
manifold giving origin to the homoclinic tangle which is a signature of chaos in the model.
\par This manifestation of homoclinic chaos is illustrated in Fig. 11 where the fractality of
the initial conditions basin boundaries, connected to the code recollapse/escape into inflation is
shown. We select the initial condition set whose projection on ($\varphi,p_{\varphi}$) is a square
of characteristic length $R \simeq 10^{-3}$, constructed about the point ($a=0.15,p_a=3.114327925239397,
\varphi=0,p_{\varphi}=0$) close\cite{euclid} to the separatrix $\mathcal{S}$ reaching $P_1$,
containing $N=160,000$ initial conditions. These points obviously
satisfy the constraint Eq. (\ref{eq27}), with $E_{\rm dust}=E_{\rm rad}=10^{-3}$ and $|\sigma|=762$,
such that $|\sigma|E_{rad}$ is slightly below $E_{\rm crit} \equiv V_{\rm max}$.
We note that these sets correspond to initial conditions for expanding universes that visit
a neighborhood of the saddle-center $P_1$ before either to recollapse or to escape into inflation.
The points of the square are color-coded according to their asymptotic behaviour for a time $\tau=4,000$ --
white corresponding to escape into inflation and black corresponding to recollapse to the neighborhood
$a \simeq 0$ -- resulting in the plots of Fig. 11 for $m=6$ and $m=18$. We notice that as we
increase $m$, namely, as the system becomes more nonintegrable, the dominance of escape into inflation
and fractality of the boundaries increase, characterizing a chaotic exit to inflation.
The dynamics involves no amplification of the inflaton field (namely $\varphi$ and/or $p_{\varphi}$),
as opposed to the metastable nonlinear resonance regime before escape to inflation, described in
Sections V-VI.
\begin{figure}
\begin{center}
\includegraphics*[width=8.5cm,height=11cm]{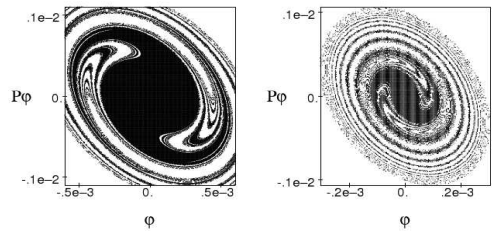}
\caption{Fractal basin boundaries, in the plane ($\varphi,p_{\varphi}$),
of initial condition sets for initially expanding universes
that visit a small neighborhood of the saddle center before either recollapsing or escaping to inflation,
in the case $E_0=E_1=10^{-3}$.
Black points correspond to orbits that recollapse to the domain $a \simeq 0$ and white points correspond
to orbits that escape to inflation, for a time $\tau=4,000$, and $m=6$ (left) and $m=18$ (right).}
\label{fig11}
\end{center}
\end{figure}
\par The entanglement and infinite transversal crossings of the stable with the unstable homoclinic cylinders
engender a further mechanism in the chaotic exit to inflation which we denote as {\it draining of initial
condition basins}. Let us remark on the fact that the surface of the cylinders constitute a boundary for the
general flow. If we start, for instance, with a set of initial conditions corresponding to initially expanding
universes, two distinct flows will be associated with these initial conditions depending on whether
they are contained inside or outside the stable cylinder. The flow corresponding to initial conditions inside
the stable cylinder will reach a neighborhood of the saddle-center $P_1$ (with $E_{a} < 0$) and will return
towards the neighborhood of $a \simeq 0$ inside the unstable cylinder, while the flow of orbits associated with
initial conditions that are outside the cylinder will reach the neighborhood of $P_1$ and escape towards the
deSitter attractor along the exterior of the unstable cylinder of the second pair. Now consider the first
transversal intersection of the cylinders: a part of the orbits inside the unstable cylinder will enter the
interior of the stable cylinder and the flow will proceed inside the stable cylinder towards the neighborhood of
$P_1$, from where it will reenter the unstable tube and proceeds towards the region $a \simeq 0$ and by a new
intersection a part of these orbits will again enter the stable tube and proceeds back towards the
neighborhood of  $P_1$ and so on. The portion of orbits that in this subsequent intersection remained outside the
stable cylinder will also proceed along it towards $P_1$ and will escape to inflation. As the motion proceeds, the successive infinite intersections of the cylinders drain the basin of initial conditions from {\it black} to
{\it white} as $\tau$ increases, as illustrated in Fig. 12, where we start from the initial condition set
of Fig. 11 for $m=6$, color-coded at $\tau=4,000$ (in the present Figures we used $N=40,000$ initial conditions only).
The only orbits remaining in an infinite
recurrence of the motion are the homoclinic orbits, which constitute the homoclinic intersection manifold,
and the infinite countable set of periodic orbits with arbitrarily long periods in the neighborhood of
each homoclinic orbit. These orbits constitute a Cantor set that is a topological characterization
of chaos in the model\cite{holmes}.
\begin{figure}
\begin{center}
\includegraphics*[width=8.5cm,height=11cm]{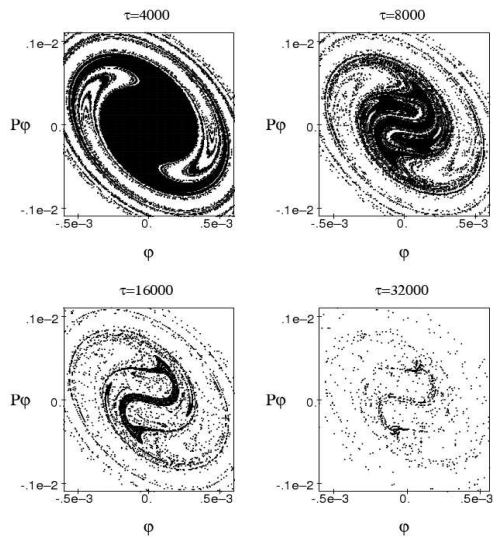}
\caption{Illustration of the process of draining (in time) the initial condition
basin of trapped orbits (black) in favor of escaping orbits (white), for $m=6$.
The initial condition set taken about the separatrix is the same as in Fig. 11.
The Figures show the almost complete exit to
inflation (up to a Cantor set of orbits) in a long time term.}
\label{fig12}
\end{center}
\end{figure}
\begin{figure}
\begin{center}
\hspace{0.0cm}
\vspace{0.0cm}
\includegraphics*[width=8.5cm,height=11cm]{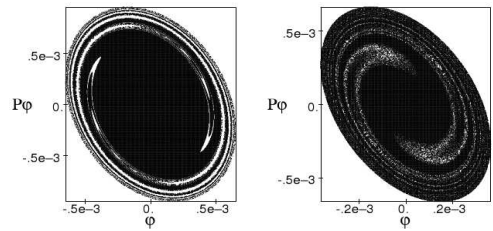}
\caption{Fractal basin boundaries in the plane ($\varphi,p_{\varphi}$) for pure dust (left) and
pure radiation (right), for $m=6$ and color-coded at $\tau=4,000$. The fractalization of the boundaries
appears less effective in both cases than in the case of radiation plus dust of Fig. 11(left).}
\label{fig13}
\end{center}
\end{figure}
\par Finally in Fig. 13 we exhibit two initial condition sets for $m=6$, color-coded at $\tau=4,000$,
in the cases of pure dust (left) and pure radiation (right), both with $N=160,000$ points.
For the radiation case the initial set was constructed about the point ($a_0=0.15$, $p_{a0}=4.144304603687830$, $\varphi_0=0=p_{\varphi 0}$), sufficiently close to the separatrix, with $|\sigma|=1499$ and $E_{\rm rad}=10^{-3}$.
For dust the initial set was constructed about the point ($a=0.15$, $p_{a}=1.460426592928701$,
$\varphi=0=p_{\varphi}$), also sufficiently close\cite{euclid} to the separatrix, with $|\sigma|=1630$
and $E_{\rm dust}=10^{-3}$, both sets also with characteristic length $R \simeq 10^{-3}$. Thus the mechanism of chaotic exit to inflation is enhanced in the case of dust plus radiation (see Fig. 11), in
comparison with the case of pure dust or pure radiation components.
\par The chaotic exit to inflation examined
in the present Section is not a feature of bouncing inflationary cosmologies only.
In fact a similar pattern for fractal
initial conditions basin boundaries connected to the escape to inflation
was observed and discussed by Cornish and Levin\cite{levin}
in the context of singular closed FRW cosmologies sourced with several conformally and/or
minimally coupled scalar fields. An extensive analysis showed that the pattern
is typical for the nonintegrable cases of the models, and
a quantitative measure of the fractality was made by evaluating
the box-counting dimension of the boundaries.
The effect of {\it draining of initial conditions} from recollapse to escape
as time increases, observed in our models, is not seen there due to
the absence of bounces that avoid the singularity and allow for a long time recurrence.
The Cantor set resulting from this effect as $\tau \rightarrow \infty$
corresponds to the strange repeller structure already appearing in their dynamics.
\section{Conclusions and Final Discussions}
In the present paper we construct nonsingular cosmological scenarios, in the realm of
string inspired braneworld models, which are past-eternal, oscillating and may emerge into
an inflationary phase due to nonlinear resonance mechanisms. We consider a closed
Friedmann-Robertson-Walker (FRW) metric on the 4-dim braneword embedded in a 5-dim conformally flat bulk.
Local bulk effects on the 4-dim FRW braneworld introduce corrections in Friedmann's
equations that allow to implement nonsingular bounces in the scale factor of the models.
This is the case of a timelike extra dimension, when the corrections result in a
repulsive force that avoids the singularity and provide a concrete model for
bounces in the early phase of the universe. The matter content of the model, confined to the
FRW brane, consists of noninteracting perfect fluids with equation of state $\rho_i=\alpha_i~ p_i$,
with $-1/3 < \alpha_i \leq 1$, plus a massive conformally coupled scalar field.
In the FRW brane the energy density of the
fluids are $\rho_i=E_i/a^{3(1+\alpha_i)}$, where the constant of motion $E_i$ is proportional
to the total energy of the respective fluid. The corrections in Friedmann's equations
resulting from the bulk-brane interaction are quadratic in the energy-momentum tensor of
the matter fields on the brane. The resulting dynamics is nonintegrable and chaotic if the mass
of the inflaton $m \neq 0$, allowing for metastable configurations that realize inflation
due to parametric nonlinear of KAM tori present in the phase space of the model.
\par Our analysis is restricted to dynamical configurations in which the dynamics of
the scale factor is initially bounded in a potential well arising in the gravitational sector
($a,p_a$) due to the effective cosmological constant, the positive spatial
curvature of the brane, and the bulk-brane corrections (connected with the perfect
fluid components) that act as infinite potential barrier and is responsible for
the avoidance of the singularity $a=0$. These configurations have the theoretical
advantage over one single bounce models in that they avoid the problem of initial conditions
for the universe at past infinity. Furthermore they are past-eternal, oscillating about a
stable Einstein universe configuration, that has no classical analogue and is favoured by
maximum entropy considerations\cite{gibbons}.
The skeleton of the phase space dynamics is analyzed
through its basic structures as critical points (stable and unstable Einstein universe),
invariant plane, separatrices and deSitter attractors at infinity, as well as
the foliation by KAM tori of the
phase space in the nonlinear neighborhood of the stable Einstein universe.
The stable Einstein universe is a critical point
of the dynamics when sourced by perfect fluids, or a limiting one-dimensional torus (namely,
the topological product of a point times a minimal periodic orbit in the scalar field sector)
when sourced by a scalar field in pure scalar field cosmologies.
\par For particular domains of the parameter space ($\Lambda_4,\sigma,E_i,m$) of the model,
denoted windows of resonance, these oscillatory bounded configurations turn metastable due to
nonlinear parametric resonance phenomena, allowing the emergence of the universe into an inflationary phase.
For numerical/analytical simplicity, we made an extensive examination of parametric resonance phenomena
in braneworld models restricted to a dark energy component and radiation ($i={\rm dust}, {\rm rad}$),
plus the massive inflaton field and a dark energy component described by the effective cosmological
constant in the brane. We constructed the resonant chart for the case of
$E_{\rm dust}=E_{\rm rad}=10^{-3}$, shown in Fig. 3,
where the windows of resonance are the gray sheets. The white regions correspond to stable motion.
When the system is driven towards a resonance window, nonlinear resonance of KAM tori takes place
resulting in a complex dynamics. KAM tori that trapped the orbits are disrupted, and the initially bounded, oscillatory orbits of the system may escape to the deSitter attractor at infinity realizing inflation.
We illustrated this behaviour by Poincar\'e maps in the plane ($\varphi^{\prime},\varphi$), the origin of
which corresponds to a stable or unstable periodic orbit whether respectively the system
is in a stable region or a resonance window of the resonance chart.
\par Each resonance window is characterized by an integer $n \geq 2$ and its main feature is
the bifurcation of the stable periodic orbit at the origin $(\varphi=0,p_{\varphi}=0)$ into
an unstable periodic orbit accompanied by one or two characteristic stable periodic orbits whether respectively $n$ is odd or even. As the initial conditions of the expectation values $\varphi$
are assumed to be small, being taken near the invariant plane $\varphi=0$, $p_{\varphi}=0$,
it follows that the parametric domains of resonance are the ones that allow for inflation
in the system. In this sense, since inflation is a sound paradigm for cosmology strongly sustained
by observations, and if our present Universe is actually a braneworld, then the values of the
cosmological parameters must be constrained to the resonance windows -- with the braneworld
inflated from initial conditions connected to a particular resonance.
In particular, for fixed $m$ and $E_i$, the brane tension $\sigma$ that regulates the strength
of the effective Newton's constant in the brane will be restricted to
small sheets depending on the integer $n \geq 2$ (cf. Fig. 3). Therefore the larger the order
of the resonance the stronger the gravitational coupling strength in the in the respective brane
inflated due to a specific resonance. In this instance, we observe a {\it quantization} of the brane
tension and consequently of the effective Newton's constant.
\par The volume of the resonance windows are small as compared to the whole volume of
the parameter space. The size of these 4-dim volumes depends strongly on the fluid content of the model.
In Fig. 5 we exhibited the windows corresponding to the $n=3$ resonance for pure dust, pure radiation
and dust plus radiation in equal amounts ($E_{\rm rad}=E_{\rm dust}$). The case in which the perfect fluid
component is dust only presents resonance windows with a reduced volume compared
with pure radiation windows, analogously to the substructure of disruptive resonances
in each window. This suggests that if we demand that our preinflationary models contain a component
of dark matter in the form of dust
this amount should be properly bounded in order that the model could realize inflation.
\par Three distinct dynamical patterns are set up by the resonance, according to
substructures in the resonance windows in models with $E_{\rm dust} \leq E_{\rm rad}$
as observed by a detailed numerical examination.
If the initial conditions correspond to configurations in the left border of the resonance
windows we have short time disruption of the initially bounded orbit with a rapid escape to
inflation. On the other hand, if the initial conditions correspond to configurations in an
intermediate threshold region of the window the orbit undergoes a long time diffusion
in the stochastic sea surrounding the two secondary KAM stability islands, with
posterior escape to inflation. Beyond the threshold region, near the right border of the window,
the orbit undergoes diffusion through large regions of phase space but remains bounded for times
larger than $\tau=100,000$. This latter region of the resonance window should then be
excluded as physically not admissable since they correspond to configurations that do not
realize inflation.
\par We also examined cosmological scenarios with a pure scalar field. Due to the absence
of any perfect fluid component, the bulk-brane correction term that allows
the dynamics to avoid the singularity is crucially dependent on the nonzero initial
amplitude of inflaton through the conserved energy $\Delta_0=({\varphi^{\prime}}(0)^2+\varphi(0)^2)/2$.
Contrary to the cases with a perfect fluid component, the stable Einstein universe is not a critical point but
a limiting 1-dim tori, sourced by a scalar field. For a fixed $\sigma$ each energy surface contains
just one torus generated by a single orbit. The escape to inflation in the nonintegrable case
is not associated to a parametric resonace pattern and occurs just for configurations
of the scalar field with large initial conditions, in the small stochastic sea on the border of
the main KAM island shown in Fig. 9. In this outer border we observe a mechanism of dynamical partial confinement of orbits, leading to finite time oscillations before escape to inflation. Also,
contrary to models with a perfect fluid component, the structure of
the bouncing dynamics is extremely sensitive to the initial amplitude
and to the mass of the scalar field, and the presence of dynamical potential
barriers allowing for bounces in the scale factor appears as a new feature of the
dynamics.
\par In Section VIII we examined the chaotic exit to inflation for initial condition sets (corresponding
to initially expanding universes) taken in a small neighborhood about the stable separatix $\mathcal{S}$.
These sets are shown to have fractal basin boundaries connected to the code recollapse/escape to inflation
leading to a chaotic exit to inflation. The fractality of the initial condition sets increases with $m$,
namely, with the nonintegrability and their escape to inflation takes place smoothly involving
no strong amplification of the inflaton field, as opposed to the metastable nonlinear resonance regime
immediately before the exit to inflation, as discussed in Sections V-VI. We also observe the phenomenum of
draining of these initial condition basins from recollapse to escape behaviour, as time increases. For
$\tau \rightarrow \infty$ only the homoclinic intersection manifold remains in recurrent oscillatory
motion. The homoclinic chaotic exit to inflation appears to be enhanced in the case of dust plus radiation,
relative to the cases with pure dust or pure radiation components, contary to what occurs in the
parametric resonance escape to inflation. We have not examined the possibility of chaotic
escape to inflation in pure scalar field cosmologies.
\par Typically variation of the parameters can shrink or stretch the resonance zones.
However the underlying pattern of resonance windows is
maintained as we have checked numerically. In this sense the pattern is said to be structurally stable.
Finally if our actual Universe is a brane inflated by a parametric resonance
mechanism triggered by the inflaton, some observable cosmological parameters
(e.g. Newton's gravitational constant on the brane and the mass of the inflaton) should then have a
signature of the particular resonance from which the brane inflated and consequently of the
particular value of the parameters ($\sigma,m, E_{\rm dust},E_{\rm rad}$)
that were favored in the early dynamical regime of the universe.
\par The authors acknowledge partial financial support from CNPQ-MCT/Brasil.
Several of the figures were generated using the \textit{Dynamics Solver} packet\cite{ds}.


\begin{thebibliography}{99}
\bibitem{correction} The correction of $\nu_{\varphi}=\frac{1}{2 \pi}$ into $\tilde{\nu}_{\varphi}$
in Eq. (\ref{eq31r}) is obtained by using the exact equation
$\ddot{\varphi}+(1+m^{2}a^{2}(\tau))\varphi=0$, substituting in it the solution
$a_0(\tau)$ of the associated integrable case and extracting the dominant correction.
For a distinct ratio of $E_{\rm rad}$ and $E_{\rm dust}$ the factor $0.9$ is changed
according to the approximate medium point of the oscillations of $a_0(\tau)$. These
approximations are very efficient to localize the resonances, as shown in Fig. 3.
\bibitem{euclid} The minimal distance from this point to the separatrix, in the Euclidean sense,
is $\delta \simeq 10^{-3}$.
\end{thebibliography}
\end{document}